\documentclass[journal]{IEEEtran}

\usepackage{cite}
\usepackage{balance}
\ifCLASSINFOpdf
\usepackage[pdftex]{graphicx}
\usepackage{epstopdf}
\usepackage{subcaption}
\usepackage{pgfplots}
\usepackage{amsmath,stmaryrd,pict2e,picture}
\usepackage{float}
\else
\fi
\usepackage{amsmath,amssymb,amsthm}
\usepackage{mathrsfs}
\usepackage{algorithm}
\usepackage{algpseudocode}

\begin{document}
\title{Sparse Array Capon  Beamformer Design Availing Deep Learning}
\author{Syed~A.~Hamza,~\IEEEmembership{ Member,~IEEE}
        and~Moeness~G.~Amin,~\IEEEmembership{Fellow,~IEEE}
\thanks{This work is supported by National Science Foundation (NSF) award no. AST-1547420. }% <-this % stops a space
\thanks{Syed A. Hamza is with the School of Engineering, Widener University, PA 19013 USA, Moeness G. Amin is with the Center for Advanced Communications (CAC),  College of Engineering, Villanova University, PA 19085-1681 USA (e-mails: shamza@widener.edu, moeness.amin@villanova.edu).}
}
\maketitle
\begin{abstract}

The paper considers sparse array design for receive beamforming achieving maximum signal-to-interference plus noise ratio (MaxSINR). We develop a design approach based on supervised neural network where class labels are  generated using an efficient  sparse beamformer  spectral analysis  (SBSA) approach. SBSA uses explicit information of the unknown narrowband interference environment for training the network and bears close performance to training using enumerations, i.e., exhaustive search which is computationally prohibitive for large arrays. The employed DNN effectively approximates the unknown mapping from the input received data spatial correlations  to the output of  sparse configuration with effective interference mitigation capability. The problem is posed as a multi-label classification problem where the selected  antenna locations achieving MaxSINR are indicated by the output layer of DNN. In addition to   evaluating the  performance of the DNN in terms of the classification accuracy, we evaluate the performance   in terms of the the ability of the classified sparse array to mitigate interference and maximize signal power.  It is shown that even in the case of miss-classification, where at least one sensor location doesn't match the optimal locations, the   DNN effectively learns the sub-optimal sparse configuration which has desirable SINR characteristics. This shows the ability of the DNN to learn the proposed optimization algorithms, hence paving the way for efficient real-time implementation.

%This approach proves important for limited aperture that constrains the number of possible uniform grid points for sensor placements. 

%The problem is formulated as quadratically constraint quadratic program (QCQP), with the cost function penalized with weighted $l_1$-norm squared of the beamformer weight vector. Simulation results are presented to show the effectiveness of the proposed algorithms for array configurability in the case of both single and general rank signal correlation matrices. Performance comparisons among the proposed sparse array, the commonly used  uniform arrays, arrays obtained by other design methods, and arrays designed without the augmentability constraint are provided.
\end{abstract}
\IEEEpeerreviewmaketitle
\section{Introduction}
Sparse array design reduces system transceiver costs  by reducing the hardware  and processing complexity through sensor selection. It is useful in multitude of sensor signal processing tasks in MIMO communications, radar/sonar, satellite navigation,  radio telescopes, speech enhancement  and medical imaging applications \cite{710573, 6477161, 1428743, 4058261,  4663892,   6031934}. The performance gains in using sparse arrays stem from their inherent ability of tending the additional degrees of freedom to accomplish  pre-defined  metrics. Several different performance metrics have been proposed in the literature, and can generally be categorized  into environment-independent or environment-dependent design \cite{AMIN20171, HAMZA2020102678}. The latter requires array reconfigurability since, the receiver performance then  largely depends on the operating environment, which may change according to the source and interference signals and locations.   This is in contrast to environment-independent sparse arrays whose configurations follow certain formulas and seek to attain structured sparse configurations with  extended aperture co-arrays. The driving objective, in this case,  is to enable direction of arrival (DOA) estimation of more sources than the available physical sensors. Common examples of structured sparse arrays are the nested and coprime arrays \cite{1139138, 5456168, 7012090}.

Reliably extracting a desired  signal waveform  by enhancing SINR has a direct bearing on improving target detection and localization for radar   signal processing,   increasing   throughput or  channel capacity for MIMO wireless communication systems, and enhancing   resolution capability in  medical imaging \cite{Goldsmith:2005:WC:993515, Trees:1992:DEM:530789, 1206680}.  Maximizing  signal-to-noise ratio (MaxSNR) and MaxSINR criteria have been shown to yield significantly efficient beamforming performance and interference  mitigation. For sparse array design, the  MaxSINR beamforming performance depends mainly on the selected positions of the sensors as well as the locations of the desired source and interferers in the field of view (FOV) \cite{6774934,  1634819, 6714077, 8036237, 8061016, 8892512}.  It is noted that with sparse arrays,  Capon beamforming must not only find the optimum weights, as commonly used in uniform arrays, but also the optimum array configuration \cite{1206680}.  This is clearly an entwined optimization problem and requires attaining maximum SINR considering all possible sparse array configurations.    \par

Sparse array design typically involves the selection of a subset of uniform grid points for sensor placements. For a given number of sensors, it is assumed that the number of perspective grid points, spaced by half wavelength, is limited due to  a size constraint on the physical aperture. For environment-dependent sparse arrays, the antenna positions are selected from uniformly spaced locations that are served by a limited number of transceiver chains. These antenna positions would vary with the changing environment. Rapid array reconfigurability has been made possible by  advances in efficient sensor switching technologies that readily activates a subset of sensors on a predefined grid points. The system cost can then be significantly  reduced by limiting the number of expensive transceivers chains at any given time \cite{738234, 967086, 922993, 1299086}.

%as shown in Fig. \ref{Block_diagram_ch8} 
Environment-dependent sparse array design algorithms generally follow two different approaches. The designs based on prior knowledge of interference parameters, essentially require that the real time interference parameters, such as DOAs and respective SINRs, are either known or  estimated apriori in real time. The other approach is more practical as it doesn’t  require the information of interfering environment which is the case in  Capon beamforming. In both cases, several iterative algorithms have been proposed to optimize the sparse array beamformer design. Although, convex based optimization algorithms, such as semidefinite relaxation (SDR) and successive convex approximation (SCA) have been developed to yield sparse configurations with desirable beamforming performances \cite{8061016, 8892512},  real time implementations of these algorithms remain limited due to the relatively high computation cost. The problem becomes more pronounced   in rapidly changing environments which result from temporal and spatial non-stationary behaviors of the  sources  in the field of view.

In this paper, we propose a sparse beamformer methodology implementing  data-driven array design by training the DNN to learn and mimic the sparse beamforming design algorithms. DNNs have shown great potential  due to their  demonstrated ability of  feature learning  in many applications, including computer vision, speech recognition, and natural language processing \cite{articlelecun1, articlelecun3, inproceedingslecun2}. In the underlying problem, DNN is used to approximate  the unknown mapping from the  receiver data spatial correlations  to the output array configuration. It is shown that DNN effectively learns the optimum array structure which makes DNN, in requiring a few
 simple operations, suitable for  real-time implementation.
%(Fig. \ref{DNN_diagram_ch8})

Towards DNN-based sparse array design, the training data may follow two different approaches. In the first approach, we use enumeration technique to generate the training labels for any given  sensor correlation function. In this case,  MaxSINR array configuration is obtained by sifting through all possible sparse configurations and choosing the best performing array topology. Although the training data is generated offline,  it becomes infeasible to obtain an  optimum configuration even for a moderate size arrays due to the enormous number of sensor permutations. In order to circumvent this problem, we propose the second approach that expedites the generation of a large number of training data labels to input the DNN. For any specific environment presented  in the training set, the proposed  approach considers the corresponding array spatial spectrum and incorporates  the sensor correlation lag redundancy for determining the desired array structure. 

%Aside from efficient generations of DNN training data, this approach can itself be used as a stand alone method to determine the best array configuration if prior information of the interference parameters is provided.   
 %sparse beamformer spectral analysis (SBSA) design algorithm that not only performs extremely well under the prior information of the interference parameters, but also expedites

%Recently, 'learn to optimize' approaches has been proposed to improve and  automate the implementation of DNNs iteslf, which  largely require  the laborious task  of designing the learning algorithms alongside  model and hyperparameter selection that needs to be manually  tailored from task to task.   In these methods, an additional DNN is trained  (called the meta-DNN) to   ensure better optimization of the original DNN (called the base-DNN)  and  generalizes  over   many tasks, hence avoiding to redesign algorithm for  closely related   tasks \cite{DBLP:conf/iclr/LiM17, 10.5555/3157382.3157543, oshea2016recurrent}. 
Recently, `learn to optimize' approaches have been proposed to improve and  automate the implementation of  learning algorithms alongside  model and hyperparameter selection that needs to be manually  tailored from task to task \cite{DBLP:conf/iclr/LiM17, 10.5555/3157382.3157543}. Depending on the task at hand, the  DNN employed   can either be  trained by reinforcement learning or supervised learning \cite{oshea2016recurrent}. It has been shown that  reinforcement learning is effective  in the case when the training samples are not independent and identically distributed (i.i.d.). This is precisely the case in optimizer learning  since the step vector towards the optimizer,  for any given iteration, affects the gradients at all subsequent iterations. 
On the other hand,  the DNN  design based on standard  supervised learning approach has been shown  to realize computationally efficient implementation of  iterative complex signal processing algorithms \cite{8444648,  DBLP:journals/twc/ShenSZL20}. In essence, the DNN learns from the training examples that are generated by running  these algorithms offline.  Efficient algorithm online implementation is then realized by  passing the input through the pre-trained DNN -- a process which only requires a small number of simple operations to yield the optimized output.

In this paper, we develop sparse array beamformer design  implementation using supervised training. Learning  sparse optimization techniques has been recently studied  in the contest of developing sparse representations and seeking simpler  models \cite{articlesparseli}. In the problem considered, the sparsity is in the sensor array rather than in the scene or the sensed environment.  Learning sparse representations, thus far, has been mainly focused   on iterative “unfolding” concept implemented by a single layer of the network \cite{DBLP:journals/corr/OSheaEC17, hershey2014deep, doi:10.1137/080716542, 10.5555/3104322.3104374, articlesparseli}. In this case, the employed approach  approximates rather simple algorithms implemented through iterative soft-thresholding such as  ISTA algorithm for sparse optimization. Sparse beamformer design, on the other hand, involves intricate  operations such as singular value decomposition and matrix inversion. Also,   these designs are mainly  implemented through convex relaxation  that  are based on SDR and SCA algorithms and use sparsity promoting regularizers, rendering them  very expensive for real-time implementations. Deep learning approaches for re-configurable arrays  and antenna selections have been studied recently for DOA estimation  and efficient beamforming for communication systems  \cite{elbir2019cognitive, elbir2019joint, elbir2019joint2, 9108299}. The overarching premise in these papers is to avoid solving a  difficult optimization problem and shifting problem complexity to the neural network training phase which may be carried out off-line.  Specifically, thinning the array for direction finding was performed in  \cite{elbir2019cognitive}  using CNN and minimum square error criterion. The problem is cast as multi-class classifications where the labels are thinned arrays, i.e., subarrays. Labeling is done based on CRB with the network input representing the estimated covariance matrix. However, \cite{elbir2019cognitive} deals with only one target and it requires the full array to be active at certain radar scans. The latter negates one of the principal reasons for employing sparse arrays, namely a limited number of front-end receivers.   The proposed Capon based methodology is applicable to multiple sources and considers leakage from other possible sources in addition to the sensor noise.  
% These algorithms adopt the sparsity promoting regularized objective functions or greedy selection type algorithms. Sparse implemtaion has been considered before, but 
% The first approach we tried was to treat the problem of learning optimizers as a standard supervised learning problem: we simply differentiate the meta-loss with respect to the parameters of the update formula and learn these parameters using standard gradient-based optimization. (We weren’t the only ones to have thought of this; (Andrychowicz et al., 2016) also used a similar approach.)
% This seemed like a natural approach, but it did not work: despite our best efforts, we could not get any optimizer trained in this manner to generalize to unseen objective functions, even though they were drawn from the same distribution that generated the objective functions used to train the optimizer. 
% types of iteration
% gradient descent 
% sdr algorithms
% greedy type algorithms
% optimization algorithms as a learning problem by  train optimizers that are tailored to particular classes of functions. 
%\subsection*{} 
\begin{figure}[!t]
	\centering
	\includegraphics[width=3.35 in, height=2.2 in]{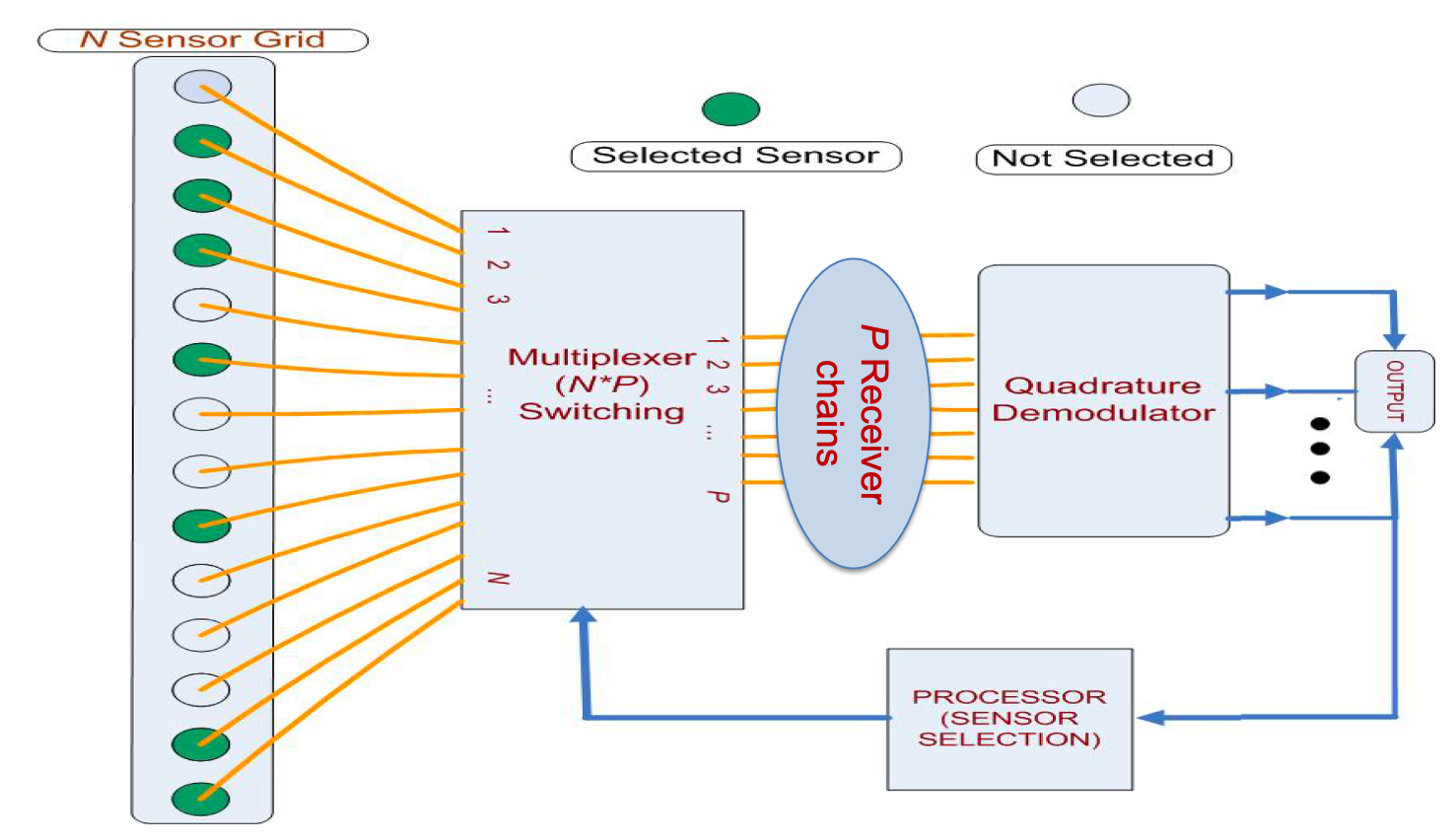}
	\caption{Block diagram of adaptive switched sensor beamformer}
	\label{Block_diagram_ch8}
\end{figure}
 
\textit{Main Contributions:} The main contributions of this paper are,

1) Sparse beamformer spectral analysis (SBSA) algorithm is proposed which is computationally efficient and  provides insights into  MaxSINR beamformer design.

%The design elucidates the concept of 'inherently amenable’ sparse configurations for interference mitigation.

 2) The DNN based approach is developed, for the first time, to configure a Capon based data driven sparse beamformer by learning the enumerated algorithm as well as SBSA design. The design is achieved through a direct mapping of the sensor data correlations  to the optimum sparse array configuration for a given `look direction'. The proposed methodology utilizes the merits of the data dependent designs, through online implementation, and   exploits the benefits of  assuming prior information of the interfering environment by efficient training with low computational complexity.
 
The proposed design is shown to be robust to limited data snapshots and can be easily  extended to robust adaptive beamforming to cater the uncertainty regarding the look direction DOA as well as array calibration errors.

The rest of the paper is organized as follows: In the next section, we state the problem formulation for maximizing the output SINR. Section III deals with the  sparse array design by SBSA algorithm and section IV describes DNN based Capon implementation. In section V, with the aid of a number of design examples, we demonstrate the usefulness of proposed algorithms in achieving MaxSINR sparse array design. Concluding remarks follow at the end.

\section{Problem Formulation} \label{Problem Formulation}
 The block digram of Fig. \ref{Block_diagram_ch8} depicts the essence of spare array beamforming.  Dark circles indicated selected antennas that are connected to the front-end receivers through a multiplexing process. Consider  a desired  source and $L$ independent interfering  sources whose signals impinge  on a  linear array with $N$ uniformly placed sensors. The baseband data received at the array  at  time instant $t$ is then given by,
\begin {equation} \label{a}
\mathbf{x}(t)=   (\alpha(t)) \mathbf{s}( \theta)  + \sum_{l=1}^{L} (\beta _l(t)) \mathbf{v}( \theta_l)  + \mathbf{n}(t),
\end {equation}
where, $\mathbf{s} ({\theta_k})$  and $\mathbf{v} ({\theta_l})$ $\in \mathbb{C}^N$ are  the  steering vectors   corresponding to the direction  of arrival, $\theta$ or $\theta_l$ of the desired source and $l$th interference, respectively, and are defined  as follows,
\vspace{+2mm}
\begin {equation}  \label{b}
\mathbf{s} ({\theta})=[1 \,  \,  \, e^{j (2 \pi / \lambda) d cos(\theta)  } \,  . \,   . \,  . \, e^{j (2 \pi / \lambda) d (N-1) cos(\theta)  }]^T.
\end {equation}
where $d$ is the inter-element spacing  and ($\alpha (t)$, $\beta _l(t))$ $\in \mathbb{C}$  are the complex amplitudes of the incoming baseband signals \cite{trove.nla.gov.au/work/15617720}. The additive Gaussian noise $\mathbf{n}(t)$ $\in \mathbb{C}^N$ has   variance   $\sigma_t^2$.
The elements of the received data vector $\mathbf{x}(t)$   are   combined linearly by the $N$-sensor  beamformer that strives to maximize the output SINR. The output signal $y(t)$ of the optimum beamformer for maximum SINR is given by \cite{1223538}, 
\begin {equation}  \label{c}
y(t) = \mathbf{w}_o^H \mathbf{x}(t),
\end {equation}
where $\mathbf{w}_o$ is the optimum weight vector resulting in the optimum output SINR$_o$,
\begin{equation}  \label{c2}
\text{SINR}_o=\frac {\mathbf{w}_o^H \mathbf{R}_s \mathbf{w}_o} { \mathbf{w}_o^H \mathbf{R}_{s^{'}} \mathbf{w}_o}.
\end{equation}
For statistically independent signals, the desired source correlation matrix is $\mathbf{R}_s= \sigma^2 \mathbf{s}( \theta)\mathbf{s}^H( \theta)$, where $ \sigma^2 =E\{\alpha (t)\alpha ^H(t)\}$. Likewise,  the  interference and noise correlation matrix, $\mathbf{R}_{s^{'}}= \sum_{l=1}^{L} (\sigma^2_l \mathbf{v}( \theta_l)\mathbf{v}^H( \theta_l)$) + $\sigma_t^2\mathbf{I}_{N\times N}$,    with $ \sigma^2_l =E\{\beta _l(t)\beta_l^H(t)\}$ being the power of the $l$th interfering source. 

     There exists a closed form solution   to maximize the SINR expression in (\ref{c2}) and  is given by $\mathbf{w}_o = \mathscr{P} \{  \mathbf{R}_{s^{'}}^{-1} \mathbf{R}_s  \}=\mathscr{P} \{  \mathbf{R_{xx}}^{-1} \mathbf{R}_s  \}$ \cite{1223538}. The operator $\mathscr{P} \{. \}$  computes the principal eigenvector of the input matrix. Substituting $\mathbf{w}_o$ into  (\ref{c2})  yields the corresponding optimum output SINR$_o$,
\begin{equation}  \label{f}
\text{SINR}_o= \Lambda_{max}\{\mathbf{R}^{-1}_{s^{'}} \mathbf{R}_s\}.
\end{equation}

\begin{figure*}[!t]
	\centering
	\includegraphics[width=6.55 in, height=2.3 in]{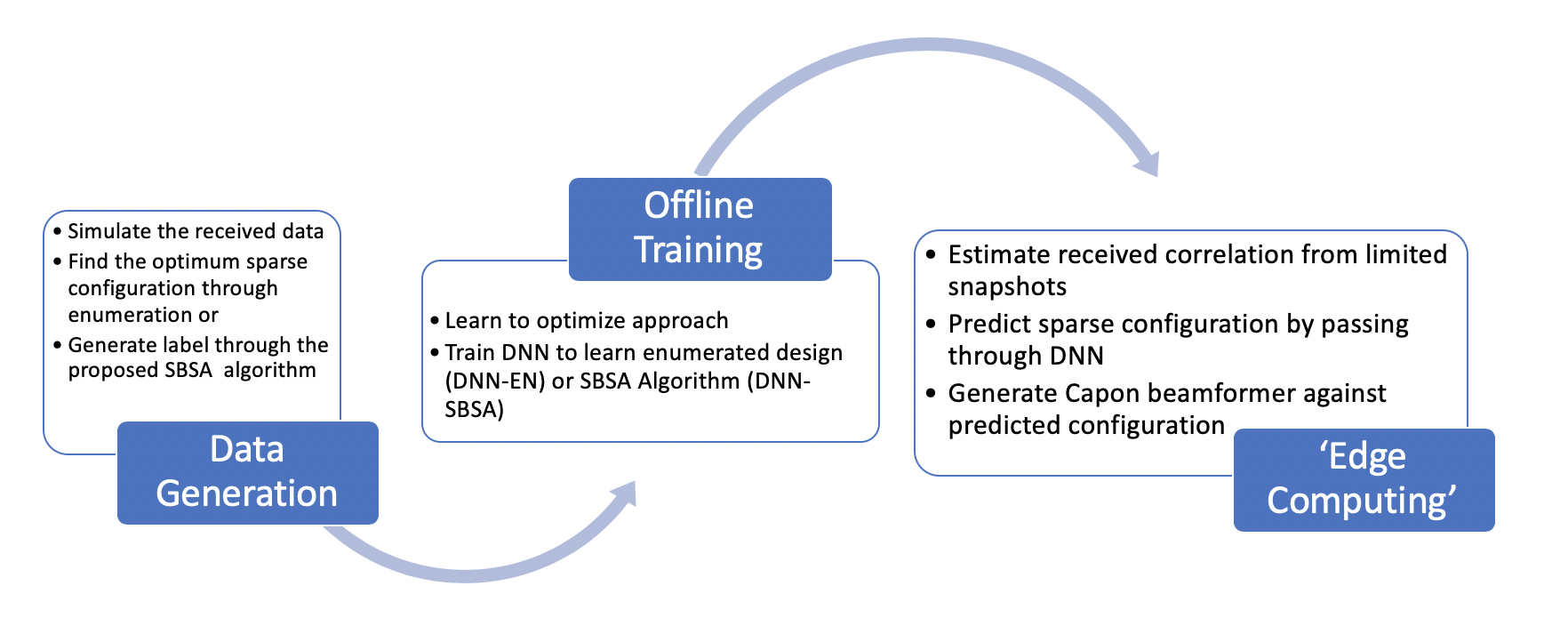}
	\caption{Overview of the proposed  approach using Deep Neural Network (DNN)}
	\label{Smart_diagram_ch8}
\end{figure*}
Accordingly,  the optimum output SINR$_o$ is given by the maximum eigenvalue ($\Lambda_{max}$) associated with   the product of  the inverse of interference plus noise correlation matrix  and the desired source correlation matrix. This is the general expression for point and distributed sources and give the similar expression in the point source case. Therefore, the performance of the optimum beamformer for maximizing the output SINR is directly related to the correlation matrix of the desired source and that of interference-plus-noise. 

\section{Sparse array design}  \label{Optimum sparse array design}
In order to maximize the SINR expression in (\ref{c2}), we constraint the numerator  and minimize the denominator term as follows,
\begin{equation} \label{d}
\begin{aligned}
\underset{\mathbf{w} \in \mathbb{C}^N}{\text{minimize}} & \quad   \mathbf{w}^H\mathbf{R}_{s^{'}}\mathbf{w},\\
\text{s.t.} & \quad     \mathbf{ w}^H\mathbf{R}_{s}\mathbf{ w}=1.
\end{aligned}
\end{equation}
The problem in (\ref{d}) can  be written equivalently by replacing $\mathbf{R}_{s^{'}}$ with the received data covariance matrix, $\mathbf{R_{xx}}=\mathbf{R}_s+ \mathbf{R}_{s^{'}}$  as follows \cite{1223538},
\begin{equation} \label{e}
\begin{aligned}
\underset{\mathbf{w} \in \mathbb{C}^N }{\text{minimize}} & \quad   \mathbf{ w}^H\mathbf{R_{xx}}\mathbf{ w},\\
\text{s.t.} & \quad     \mathbf{ w}^H\mathbf{R}_{s}\mathbf{ w} \geq 1.
\end{aligned}
\end{equation}

It is noted that the  equality constraint in (\ref{d}) is  relaxed in (\ref{e}) due to  the inclusion of  the constraint as part of the objective function, and as such, (\ref{e}) converges to  the equality constraint. Additionally, the optimal solution in (\ref{e}) is invariant up to  uncertainty in  the absolute power of the desired source.   In practice, the actual source parameters  can deviate from the  perceived ones. This discrepancy is typically mitigated, to an extent,  by pre-processing  the received data correlation matrix through diagonal loading or tapering the  correlation matrix \cite{1206680}.
 In order to bring aperture sparsity into optimum beamformer design, the  constraint optimization (\ref{e}) can be re-formulated  by incorporating  an additional constraint on the cardinality of  the weight vector;   
\begin{equation} \label{a2}
\begin{aligned}
\underset{\mathbf{w \in \mathbb{C}}^{N}}{\text{minimize}} & \quad  \mathbf{ w}^H\mathbf{R_{xx}}\mathbf{w},\\
\text{s.t.} & \quad   \mathbf{ w}^H\mathbf{R}_{s}\mathbf{ w}\geq1,  \\
& \quad   ||\mathbf{w}||_0=P.
\end{aligned}
\end{equation}
Here,  $||.||_0$ determines the cardinality of the weight   vector $\mathbf{w}$.  This is a combinatorial optimization problem and can be solved  by enumerating over all possible sensor locations.   Several different approaches have been developed to mitigate the computational expense of the   combinatorial search by either  exploiting prior given  information on the interference parameters, such as respective DOAs,   or employing  data-dependent algorithms realized through the SDR and SCA algorithms \cite{6774934, 8892512}.  These algorithms, however, have high computational costs, impeding  real time implementations, especially in applications involving rapidly changing environments. In the context of designing sparse arrays using machine learning, the optimum sparse arrays for different operating environments constitute labels for the network training data. These labels can be generated off-line  through enumerations, accounting for all possible source and interference scenarios.  For large array, however, this approach becomes computationally challenging.   In order to mitigate this  problem, we propose an efficient technique to generate DNN training samples, using   sparse beamformer spectral analysis (SBSA) design.  This technique,   detailed below,  employs what is referred to as sparse beamformer spectral
analysis (SBSA) design. It has desired  performance  and low computational complexity. The block diagram of Fig. \ref{Smart_diagram_ch8} describes the three main steps in DNN-based sparse array design using training data generated by the SBSA and enumeration.

%as compared to the state-of-the-art. The subsequent subsection presents  the implementation of the SBSA and enumerated design algorithms  using the DNN approach, this further reduces the implementation time and also paves the way to data dependent design (refer to Fig. \ref{Smart_diagram_ch8} for an overview). 
  
\subsection*{The Role of Spare Configuration in MaxSINR}
 
 As implied by   (\ref{f}), the   optimum sparse array for MaxSINR design depends on both the beamforming weights and  the  sparse array configuration.  Therefore, the problem of interference mitigation is not  a cascade design of the optimum configuration followed by the  optimum weights, or vice versa. Rather, it is  an entwined task, in essence,  calling for  joint optimization of the beamforming weights and  sparse array configuration. Albeit not optimum, the cascade design approach is not entirely without merits. It can offer a unique insight into the problem and address the suitability of a given sparse array configuration  in canceling the interfering signals.
 
 In order to shed more light on the role of array configuration in  optimum sparse beamforming, we recognize that the problem formulation developed in the previous section is valid irrespective of the array configuration and holds true for the compact ULA or any  generic sparse configuration. The beamformer output signal, $y(t) = {\mathbf{w}}^H \mathbf{x}$,  for a sparse beamformer can also be written as $\overset{\circ}{y}(t) = \overset{\circ}{\mathbf{w}}^H \mathbf{x}$, where the typeset `${{\circ}}$' indicates that the corresponding vector is sparse with few zero entries. The sparse beamformer $\overset{\circ}{y}(t)$ can  be rewritten, equivalently, as $\overset{\circ}{y}(t) = \overset{\circ}{\mathbf{w}}^H \{\overset{\circ}{\mathbf{z}}\odot \mathbf{x}\}$. The entries of $\overset{\circ}{\mathbf{z}} \in \mathbb{R}^N$ are either 1's or 0's depending on whether the corresponding sensor location is active or inactive respectively. The point-wise multiplication ($\odot$) of  the   received vector $\mathbf{x}$  with a sparse selection vector $\overset{\circ}{\mathbf{z}}$ sets the corresponding entries of the received signal to zero which amounts to the  zero beamforming weights of $\overset{\circ}{\mathbf{w}}$. 
 \begin{figure*}[!t]
	\centering
	\includegraphics[width=5.48 in, height=0.55 in]{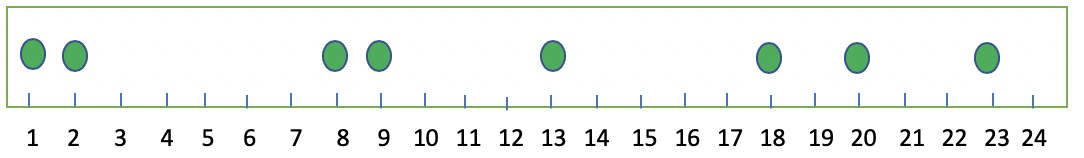}
	\caption{Eight element sparse array configuration}
	\label{mvrk-1}
\end{figure*}
\begin{figure*}[!t]
	\centering
	\includegraphics[width=6.48 in, height=2.55 in]{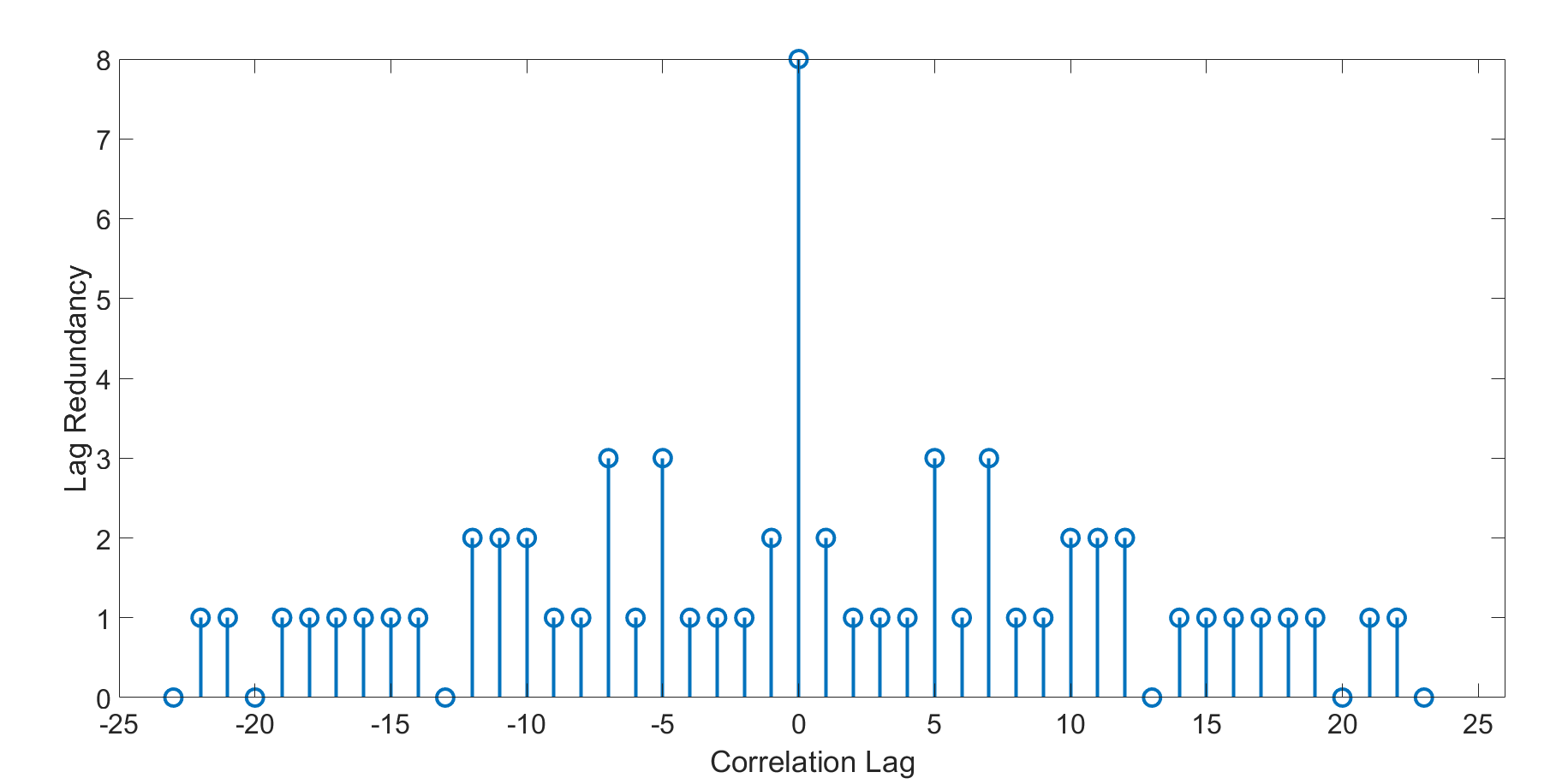}
	\caption{Lag Redundancy  of the sparse array shown in Fig. \ref{mvrk-1}}
	\label{mvrk-2}
\end{figure*}
\begin{figure*}[!t]
	\centering
	\includegraphics[width=6.48 in, height=2.55 in]{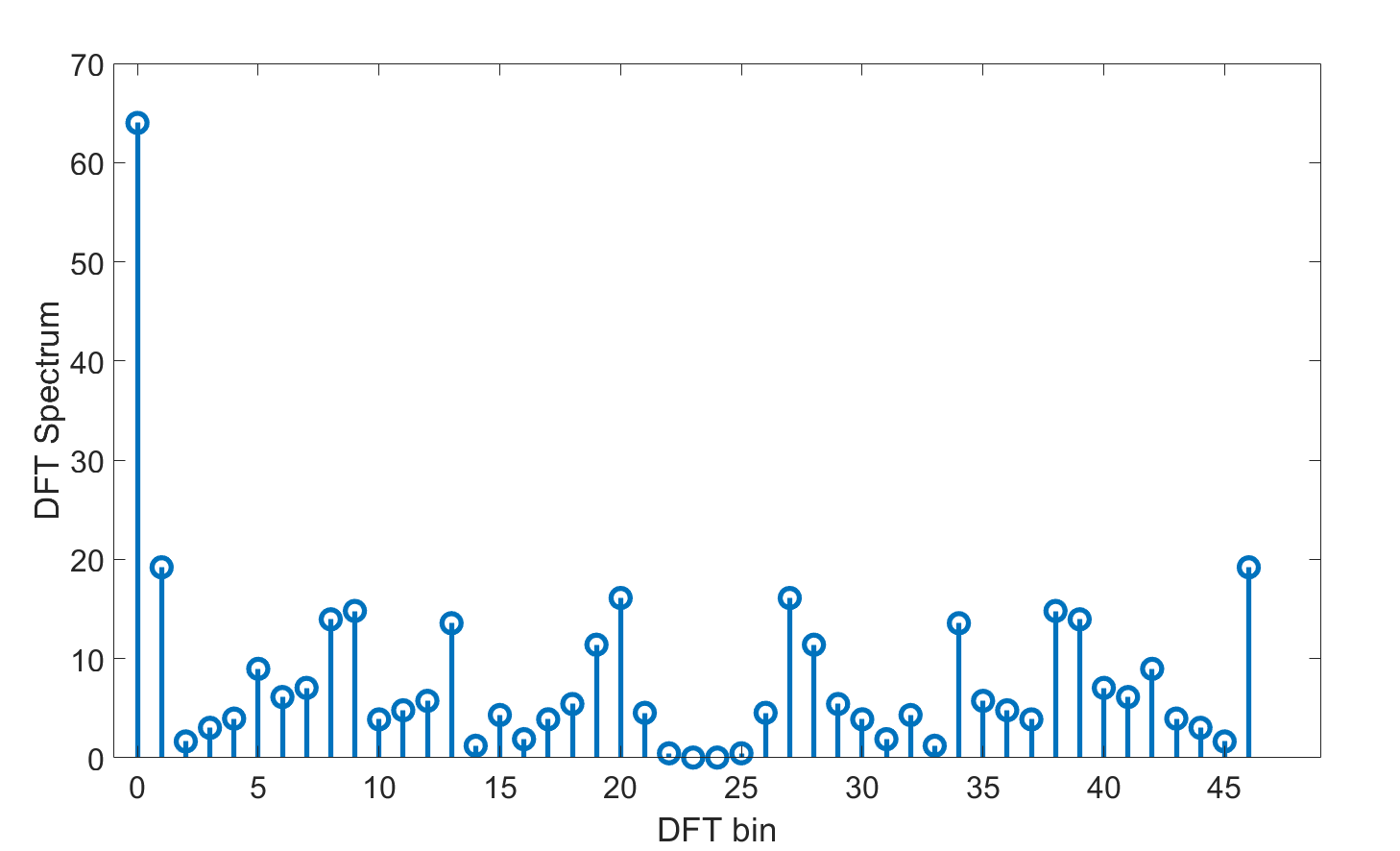}
	\caption{DFT  of the lag redundancy }
	\label{mvrk-3}
\end{figure*}
\begin{figure*}[!t]
	\centering
	\includegraphics[width=6.28 in, height=2.55 in]{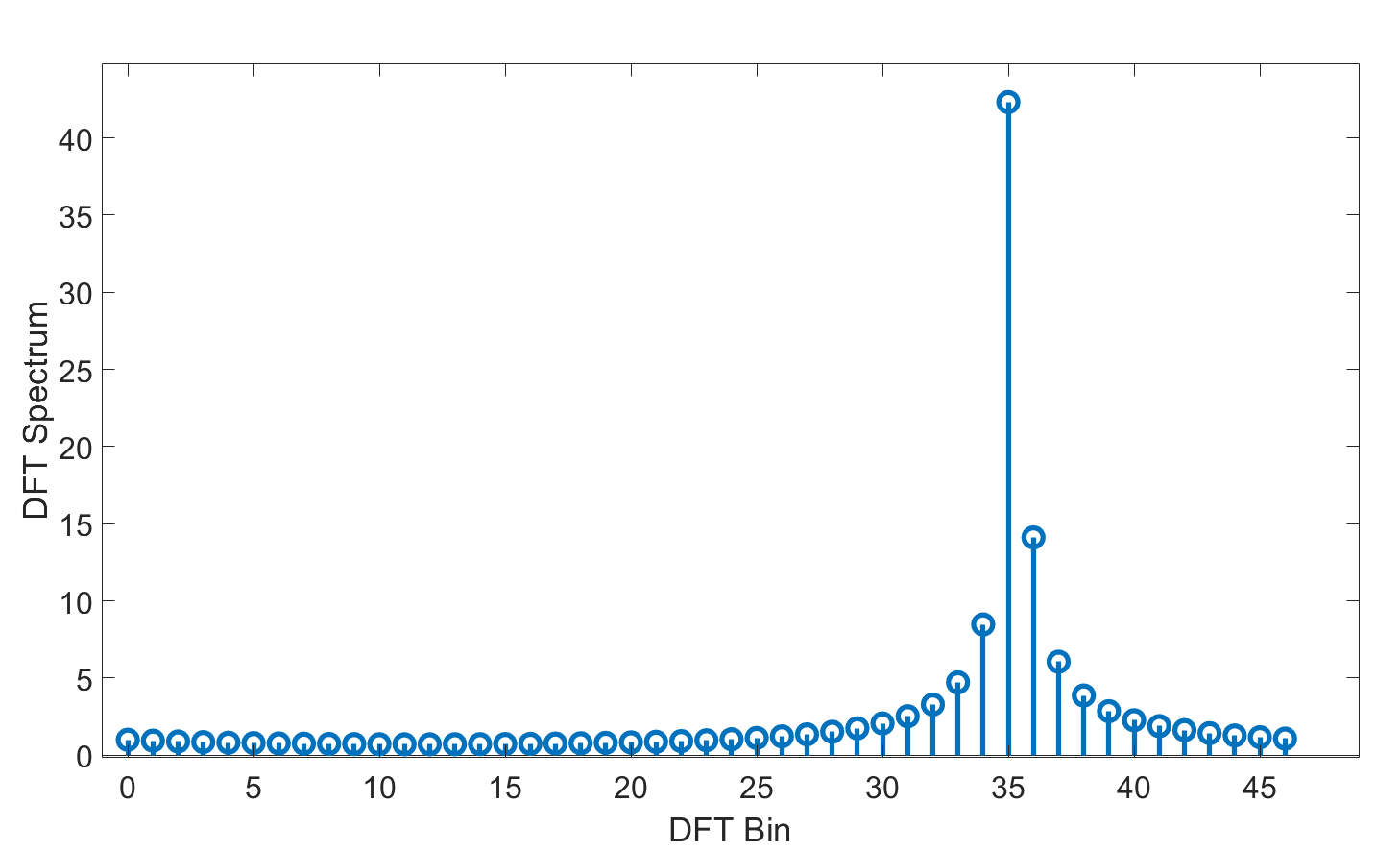}
	\caption{Power spectrum  of the desired signal}
	\label{mvrk-4}
\end{figure*}
\begin{figure*}[!t]
	\centering
	\includegraphics[width=6.78 in, height=2.75 in]{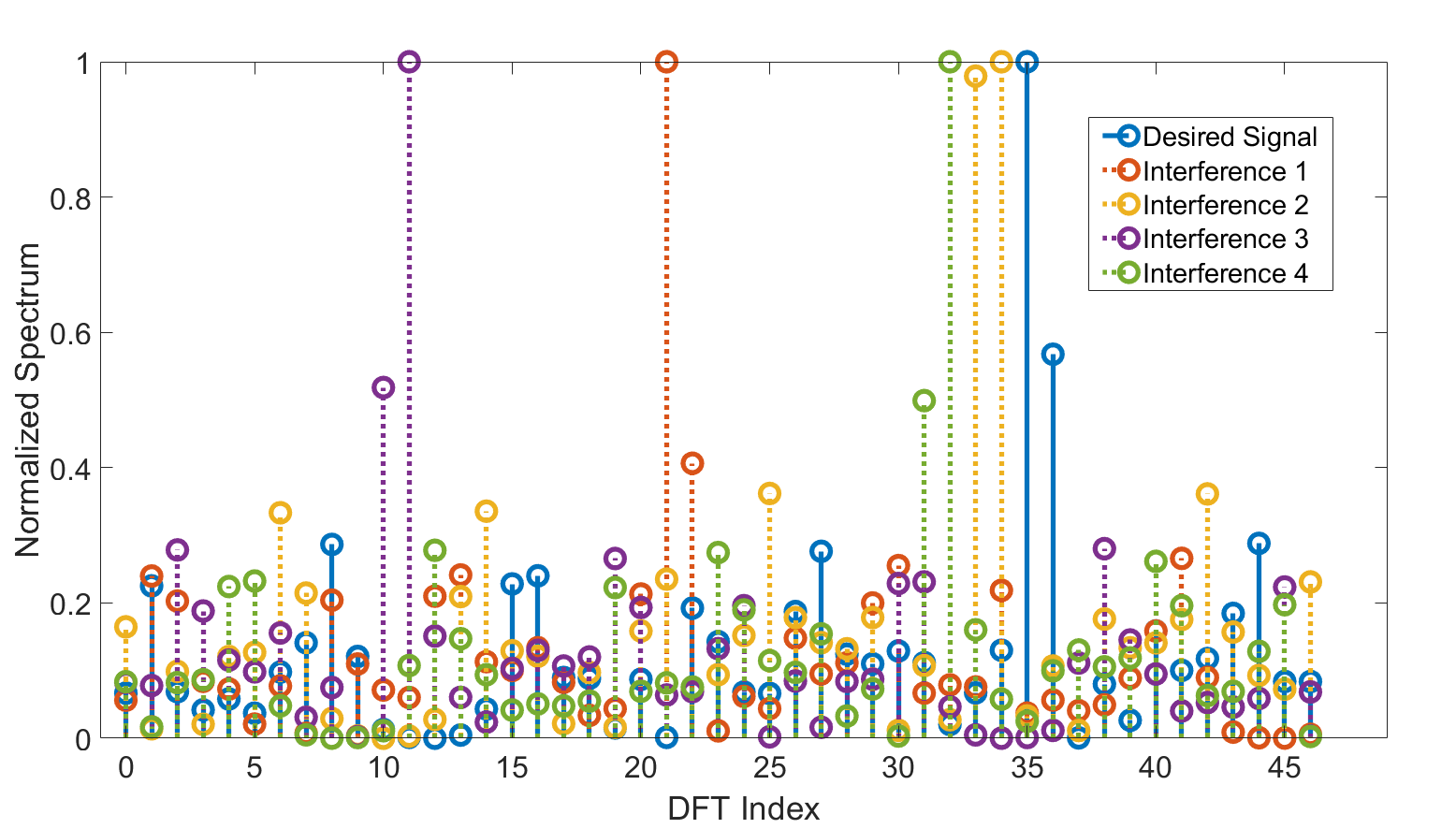}
	\caption{Explanation of the proposed objective criterion  for the optimum array configuration shown in Fig. \ref{mvrk-1}}
	\label{mvrk-5}
\end{figure*}
\begin{figure*}[!t]
	\centering
	\includegraphics[width=6.48 in, height=2.75 in]{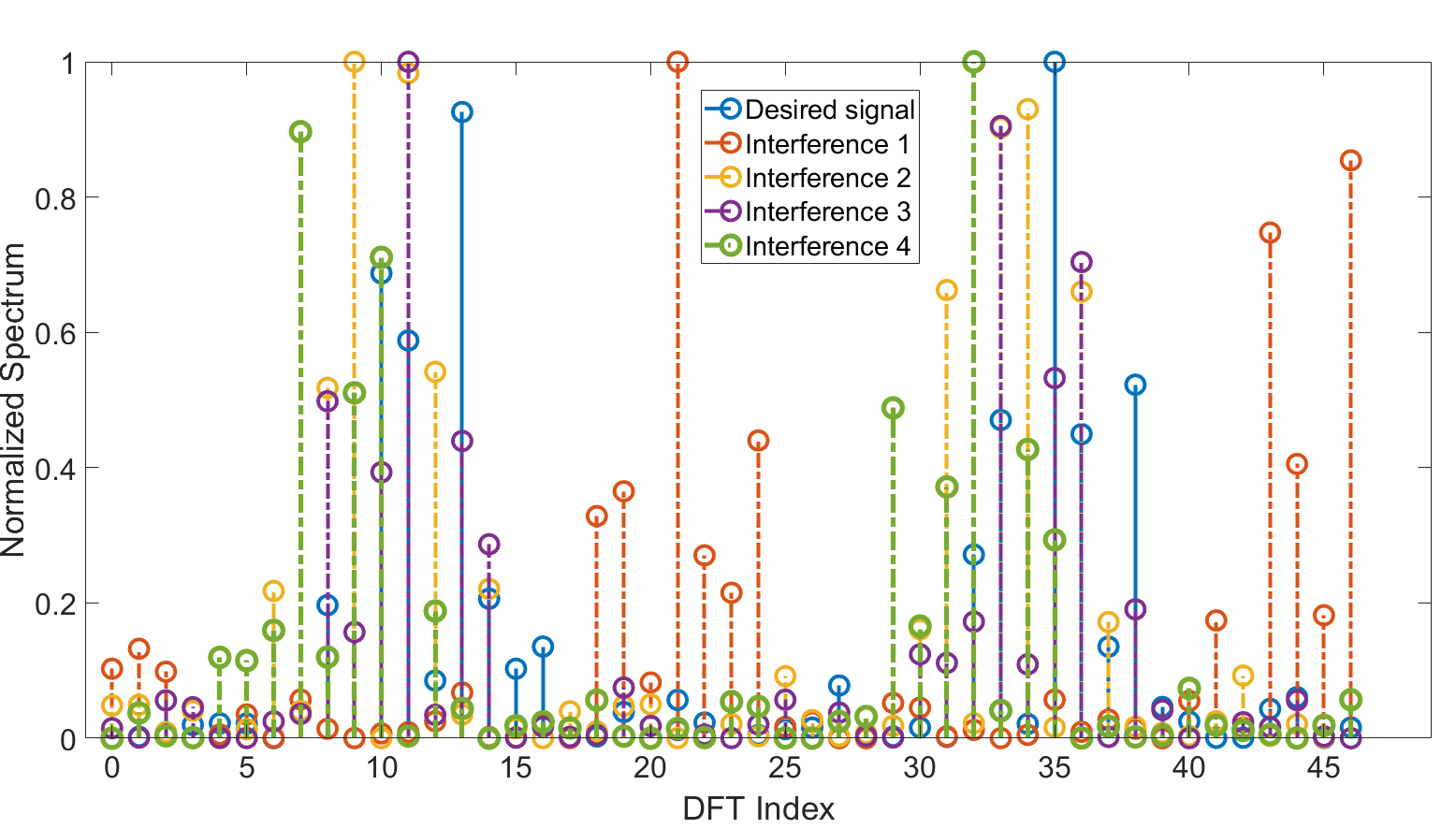}
	\caption{ Explanation of the proposed objective criterion for the worst possible array configuration}
	\label{mvrk-6}
\end{figure*}
\begin{figure*}[!t]
	\centering
	\includegraphics[width=6.48 in, height=3.15 in]{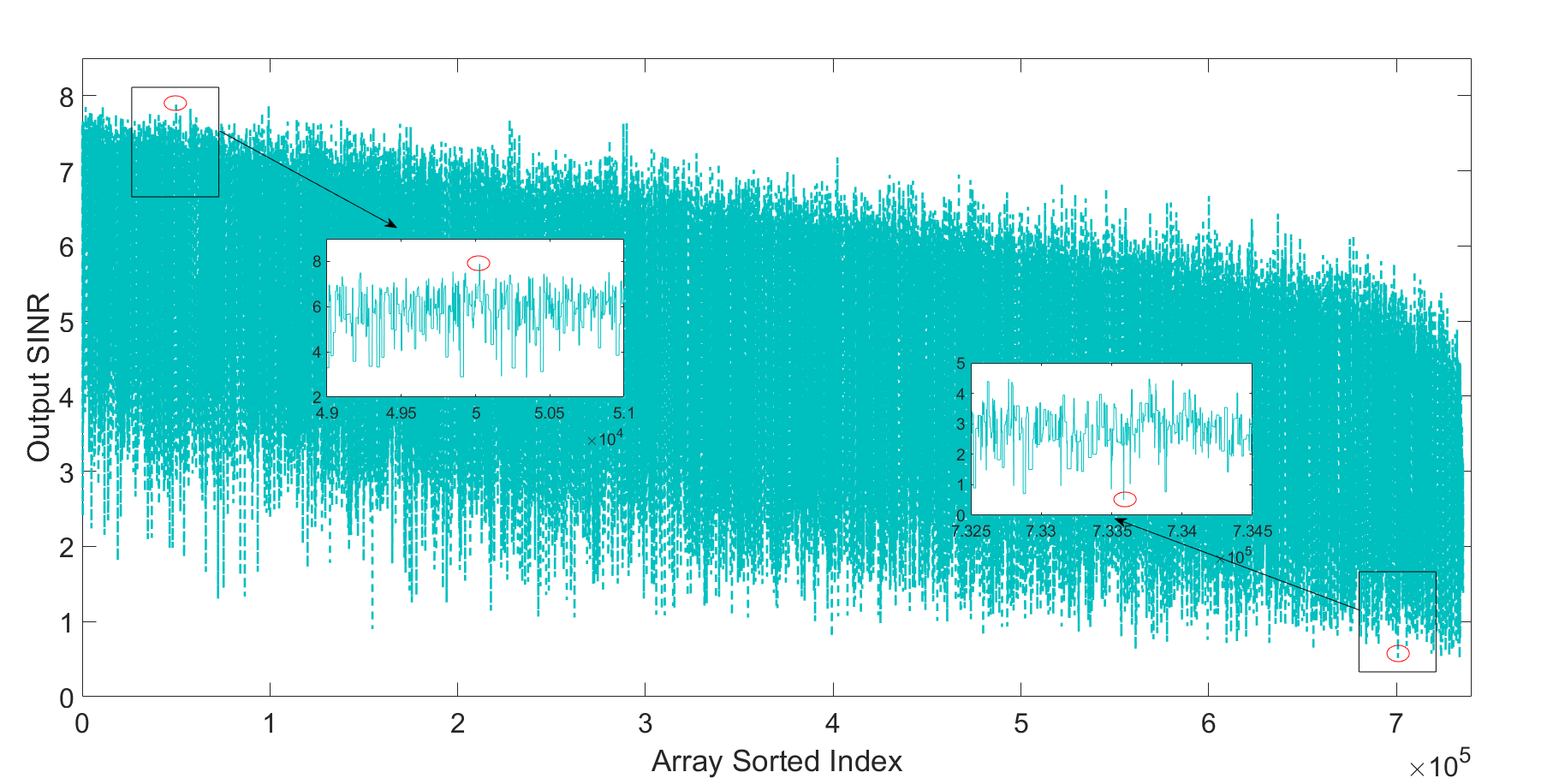}
	\caption{Plot of the proposed objective criterion   in ascending order}
	\label{mvrk-7}
\end{figure*}
 %In so doing, we analyze  the problem in the frequency domain to understand the  reshaping of the frequency content of the input signal in processing the MaxSINR sparse beamformer. 
Therefore,  sparse beamforming 
 %$\overset{\circ}{\mathbf{w}}
 can  be viewed as pre-processing the  received signal by point-wise multiplication,   prior to applying the optimum  weights. We analyze the impact of such multiplication  on the  performance of the subsequent, or cascade, design of the beamformer weights,  $\overset{\circ}{\mathbf{w}}$.  
 
 Let $\mathbf{X}= \mathscr{F}(\mathbf{x})$ and $\mathbf{Z}= \mathscr{F}(\overset{\circ}{\mathbf{z}})$ denote the DFT of the input data and the selection vector, respectively. The DFT of the point-wise multiplication of these two vectors is given by the circular convolution of the corresponding DFTs, i.e.,  $\mathscr{F}(\overset{\circ}{\mathbf{z}}\odot \mathbf{x})=\mathbf{X} \circledast \mathbf{Z}$, where  $\circledast$ denotes  circular convolution.  From (\ref{a}) and the linearity property of DFT, we obtain  $\mathscr{F}(\overset{\circ}{\mathbf{z}}\odot \mathbf{x})=\alpha \mathbf{S} \circledast \mathbf{Z}+ \sum_{l=1}^{L}\beta_l \mathbf{V}_l \circledast \mathbf{Z}+\mathbf{N}  \circledast \mathbf{Z}$, where $\mathbf{V}_l$ is the DFT of the $l$th interference vector.  The  $N$ components of $\mathbf{X}$ represent orthogonal beamformers, each pointing towards a certain spatial direction. The convolution with $\mathbf{Z}$ alters  these components as well as changes the contribution of the $l$th interference to the signal received from the desired source location.  The proposed scheme gauges the overlap between the spatial spectra of the desired signal and the interfering signals as a function of the underlying sparse configuration. A desirable array configuration would minimize such overlapping, enabling the  beamformer weights to effectively remove the interference, while maintaining the desired signal. 
 %The DFT operation is tantamount to analyzing the power distribution of the source and interfering signals by orthogonal beamformers which corresponds to DFT spatial frequencies.
 Towards this end, we propose  a design metric $\Omega(\mathbf{Z})$ based on  weighted sum of the   spatial spectrum (denoted by $|.|^2$) of the individual interfering sources scaled by the desired signal  spatial spectrum as follows,
 \begin {equation}  \label{b2}
\Omega(\mathbf{Z}) = \sum_{l=1}^{L} \{\alpha^2 |\mathbf{S} \circledast \mathbf{Z}|^2\} \odot\{\beta_l^2 |\mathbf{V}_l \circledast \mathbf{Z}|^2\}
\end {equation}
 Eq. (\ref{b2}) performs  element-wise scaling of  the interfering powers in the DFT domain.  Therefore, if the interfering signal power,  after convolution is concentrated in the    DFT bins different from   those  occupied primarily by the desired signal, then the point-wise product would assume low values. Conversely, if there is a significant overlap between the results of the two convolutions, then the objective function  $\Omega(\mathbf{Z})$ is significantly higher.
 %It is noted that  the above objective function $\Omega(\mathbf{Z})$ relies on the DFT power spectrum of the input signals and the selection vector  instead of the absolute values of the DFT. 
 
 The spatial spectrum can be estimated  by computing the  DFT of the autocorrelation function of the corresponding signal. It is worth noting that, for a given sparse configuration, the autocorrelation function of the selection vector assumes a specific  redundancy  of the autocorrelation lags. Therefore, unlike the structured sparse array design that seeks to maximize the contiguous correlation lags,  MaxSINR sparse design is guided by    the DFT of the autocorrelation sequence of the  lag redundancy.

 We   illustrate the proposed approach with the help of the following example.
  Consider an 8-element sparse array on the 24 point   equally spaced grid locations that are potentially available for sensor selections. The minimum spacing  among the  sensors is $d=\lambda/2$. Consider a source signal  located at $60^0$ and  four unwanted interferences  located at $154^0$, $55^0$, $117^0$ and $50^0$ with the  INRs ranging from $10-20$ dB. The sparse array configuration achieving the best SINR performance is found through enumeration and shown in  Fig. \ref{mvrk-1}. The associated correlation lag redundancy of this configuration and the corresponding spatial spectrum are depicted in Figs.  \ref{mvrk-2} and \ref{mvrk-3}, respectively. The spatial spectrum  of the desired source at $60^0$  prior to convolution is depicted in the Fig. \ref{mvrk-4}. The  normalized result of the two convolved  spectra in Figs. \ref{mvrk-3}, \ref{mvrk-4} is  shown as  solid lines (blue) in the Fig. \ref{mvrk-5}. The normalized  spectrum for each interfering signal is also shown as dotted lines in the same figure. Fig. \ref{mvrk-6} plots the normalized spectrum for the sparse array worst case scenario for comparison purposes. Note that the maximum of the convolved desired signal spectrum  occurs at 35th DFT bin. For the best case scenario,  all  convolved interfering signals assume minimum power at the aforementioned DFT position. This is in contrast to the  worst case scenario  where  there is  considerable interference power at the peak of the desired source. Apart form the maximum location, it is noted that for the best case design, there is  minimum overlapping between the desired signal and interfering signals at the   DFT bin locations which is clearly in contrast to the worst case design.
 
 In order to further understand the offerings  of the proposed approach, Fig. (\ref{mvrk-7}) plots the SINR performance of all  possible sparse configurations  after sorting the array configurations in ascending order  of the output of the proposed objective function of (\ref{b2}). It is clear that the average SINR in the plot is  higher and more desirable towards the left side where the objective function is minimum. It is important to note that  the best enumerated result of  MaxSINR performance does not correspond to  the array configuration with the smallest objective function. This is because, the optimum sparse configuration also depends on  the  beamformer weights to  minimize the interfering signals. This is also clear from the high variance of the curve. Similarly, the worst performing array is very close to the right side of the curve where the proposed objective function is high due to the strong overlap of the desired and undesired signal spectra.
 %The significantly reduced  performance  towards the right side of the plot can   now be  explained  intuitively  in terms of the inherent inability of these configurations to mitigate the significant overlap in the desired and interfering signals which is an artifact of the sparse beamformer explained in the frequency domain. 

 %In  light of the above discussions, it becomes prudent to  minimize the proposed objective (8) for MaxSINR design.   Towards this end, we propose
 For efficient generations of DNN training data and in lieu of enumerations, the objective function (\ref{b2}) can be minimized through 
 an iterative algorithm that implements successive sensor selections, hence deciding on  one sensor  location at a time.  For the  initial iteration, a  sensor location is chosen randomly on an $N$ grid points.  For  the $i$th subsequent iteration, the proposed objective is evaluated at the remaining $N-i$ locations and then selecting  the  sensor location that yields the minimum objective function. The procedure is iterated $P$ times for selecting the  $P$ locations from $N$ possible locations.  Due to  high variance of the curve  in Fig. \ref{DNN_diagram_ch8}, it is best to initialize  the algorithm with  different sensor location  and find the corresponding configurations for each initialization, and eventually consider the one with the  best SINR performance.  The steps of the above algorithm are detailed in  Algorithm 1.
It is noted that aside from the efficient generations of DNN training data  in the underlying problem, this approach can itself be used in general  as a stand alone method to determine appropriate  array configurations if prior information of the interference parameters is provided.   
\begin{algorithm}[t!] 
% 	\begin{table}[t!] 
 %\noindent\rule{9cm}{2.9pt}
 	\caption{SBSA Algorithm}
%	\begin{algorithmic}
		
		%\renewcommand{\algorithmicrequire}{\textbf{Input:}}
		
		%\renewcommand{\algorithmicensure}{\textbf{Output:}}
		
		\textbf{Input:}  $N$, $P$, Look direction DOA $\theta$, Interference DOAs, SNR and INRs  \\
		\textbf{Output:} Sparse beamformer $\mathbf{w}_o$ \\
				        \texttt{Initialize  $\mathbf{z}$=$[0 \, \, 1 \, \, 0\,  ... \, 0]$ where  all entries of $\mathbf{z}$ are zero except an arbitrarily selected  entry}. \\
				      \texttt{Compute the spatial spectrum of desired source and interfering signals}. \\
\textbf{for}\, \, {\texttt{j=1 to $P$-1}}\\
		\textbf{for} \, \,{\texttt{i=1 to $P$-1-j}}\\
        	  \texttt{Select the i\textit{th} sensor  from $P$-1-j remaining locations}.\\
	%	   		 \For {j=1:K-1}
    \texttt{Compute the lag redundancy of this  sparse array consisting of  j+1  sensor}.\\
   	%	   		 \For {j=1:K-1}
    \texttt{Compute the spatial spectrum of the j+1 sensor sparse array}.\\
     \texttt{Convolve the spatial spectrum of the j+1 sensor sparse array with the spectrum of the desired source and the interfering sources}.\\
      \texttt{Compute the overlapping power in the spatial spectra by computing the proposed metric in  (\ref{b2})}.\\
			      \textbf{end for}\\
		      \texttt{Select the $i$th sensor  from the inner for loop which  results in the minimum overlapping power computed by (\ref{b2})}.  \\
 \texttt{Update $\mathbf{z}$  by setting the j\textit{th}  location in $\mathbf{z}$ to 1}.\\ 
		         \textbf{end for}\\
	%	\ENDFOR
		 After finding the sparse configuration find $\mathbf{w}_o$ by running (\ref{e}) for reduced size correlation matrix while ignoring the sensor locations corresponding to $\mathbf{z}$. 
%		\RETURN $\mathbf{w}_o$
%	\end{algorithmic}
%	\noindent\rule{9cm}{2.9pt}
\end{algorithm}

\section{DNN based learning of the SBSA and enumerated  designs }

Modeling  the behaviour of optimization algorithms by training   a DNN is justifiable,  as DNNs are universal  approximators for arbitrary continuous functions and have the required capacity to accurately map almost  any  input/output formula. 
 For effective learning, it is important that the DNN can generalize to a  broader class than that represented by the finite number of drawn training examples. From the Capon beamforming perspective,   a given arrangement of a desired source direction, interference DOAs and respective SNR/INRs constitute just one particular example. A class, in this case, is defined by any arbitrary permutation of the interference DOAs and respective powers while keeping the desired  source DOA fixed.  The DNN task is, therefore, to learn from a data-set, characterized by a set of different training examples and corresponding  optimum sparse array predictions. 

Here, an important question arises as whether we could aim for a  stronger notion of generalization, i.e., instead of training for a fixed desired source, is it possible to  generalize over all possible desired source DOAs. To answer this query, we note that  for a given desired  source and interference setting, the received array data, and thereby the data correlation,   remains the same even if we switch the desired signal and one of the interferers. However, Capon beamformer should yield a different configuration due to changing  its directional constraint.   Therefore, instead of relying entirely on the information of the received data correlation, it is imperative to incorporate the knowledge of the desired source or look direction,   which is always assumed in Capon beamforming formulation. For DNN learning, this information can either be incorporated by exclusively training the DNN for each desired source DOA or the desired source DOA can be incorporated as an additional input feature to DNN. In this paper, we adopt the former approach.

\begin{figure}[!t]
	\centering
	\includegraphics[width=3.55 in, height=2.3 in]{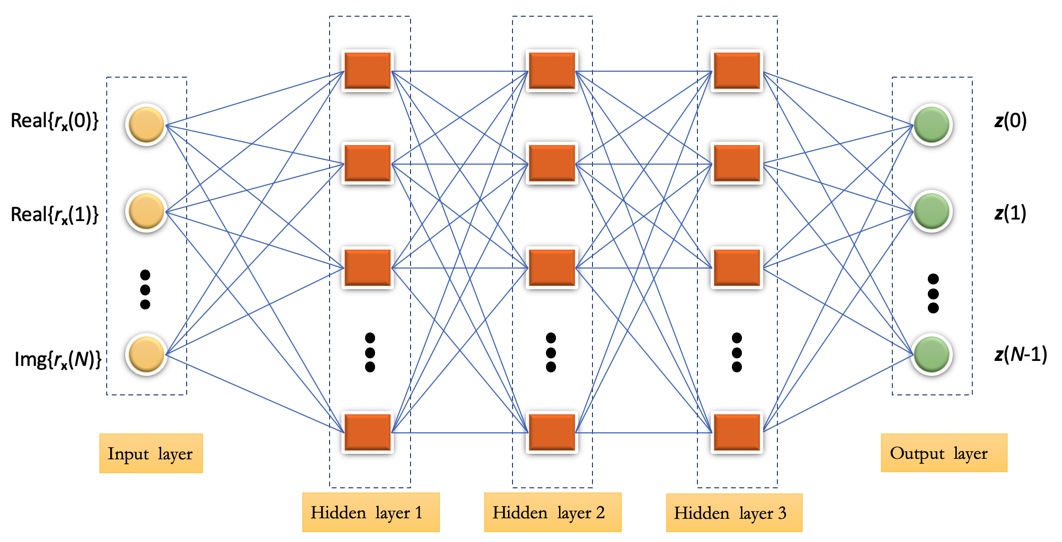}
	\caption{Architecture of Deep Neural Network (DNN)}
	\label{DNN_diagram_ch8}
\end{figure}

 For  DNN, we use multilayer perceptron (MLP) network, as shown in  Fig. \ref{DNN_diagram_ch8}. The input layer is of size $2N-1$ and the output layer is of size $N$. Although there are $N$ unique correlation lags ($\mathbf{r_x}(n-1)=\mathbf{R_{xx}}(1,n)$) corresponding to the $N$ sensor locations on the grid,  the dimensionality of the input layer is $2N-1$ owing to concatenating the real and imaginary entries of the generally complex valued correlation lags, except the  zeroth lag.  We use 3 hidden layers with 450, 250 and 80 nodes, respectively. The ReLU activation function is used for all  hidden layers activation. The  correlation  values of the received data assuming a stationary environment are input to the network. The network output $\mathbf{z}$ is a binary vector such that 1 indicate sensor selection and 0 indicates the absence of the corresponding sensor location.

The received data is generated in the following manner. For a given desired source location, the $i$th training  realization   is simulated by randomly selecting $L$ interfering signals from a DOA grid spanning the range of  $10^0$  to $170^0$, with a grid spacing of $1^0$. The interferers are allocated random powers uniformly distributed with INR from 10 dB to 20 dB. For this given scenario, the  received correlation function, which includes the desired source signal, is calculated corresponding to the full sensor configuration. The corresponding optimum configuration is found through  enumeration and also through the proposed SBSA algorithm.  The process is repeated  30000 times against a given desired source DOA to generate the training data set.  Similarly,  a small sized validation data set is generated for hyperparamter tuning for model selection, minibatch size and learning rate. 

For the training stage,  the   weights of the neural network are optimized to minimize  the mean squared error between the label and the output of the network. The ADAM algorithm is used to carry out parameter optimization, employing  efficient implementation of
mini-batch stochastic gradient descent  algorithm \cite{DBLP:journals/corr/KingmaB14}. The  learning rate is set to 0.001 and dropout regularization with the keep probability of 0.9 is used. The  weights are initialized using the Xavier initialization. We chose MLP as DNN for its simplicity.   In essence, MLP presents the baseline with other networks, such as Convolutional Neural Networks, are slated to give better performance with sufficiently large training data \cite{554195, GU2018354}.

The robustness of the learned models is demonstrated by  generating the test data  that is different from the training stage,  by  assuming the DOA of the interfering signals off grid. This is simulated by adding the Gaussian noise to the interference DOAs on the grid. We also present  the results under  limited and unlimited data snapshots. The sparse array design can only have few active sensors at a time, in essence,  making it difficult to  furnish the correlation values corresponding to the inactive sensor locations. However, for the scope of this paper, we  assume that an estimate of all the  correlation lags corresponding to the  full  aperture array are available to input for prediction. This can typically be achieved by employing  a  low rank matrix completion  strategy that permits the interpolation of the missing correlation lags \cite{HAMZA2020102678}. Additionally, to ensure the selection of $P$ antenna locations   at the output of DNN, we  declare the $P$ highest values in the output as the selected sensor locations. Therefore, generalization in this context means that the learned DNN works on different interference settings which can change according to changing environment conditions.

% \begin{figure}[t]
% 	\centering
% 	\includegraphics[width=3.38 in, height=2.5 in]{revision_16_8_symm_scan2.eps}
% 	\caption{Output SINR for different array topologies}
% 	\label{scan}
% \end{figure}
\section{Simulations} \label{Simulations}
In this section, we show the effectiveness of the proposed approach for sparse array design achieving  MaxSINR. The   results are examined first by training the DNN to learn  the enumerated optimum array configurations. Then,  we  demonstrate in the follow-on examples  the effectiveness  of the DNN when trained by the labels drawn from SBSA algorithm. 

\subsection{Enumerated design} \label{Single point source}

In this example, we pose the problem as selecting $P=6$ antennas from $N=12$  possible equally spaced locations with  inter-element spacing of $\lambda/2$. For all  numerical results, we use a network with three hidden layers, one input layer, and one output layer. Accordingly, the input to the network is of size 23 and output is size 12.

\begin{figure}[!t]
	\centering
	\includegraphics[width=3.6 in, height=2.6 in]{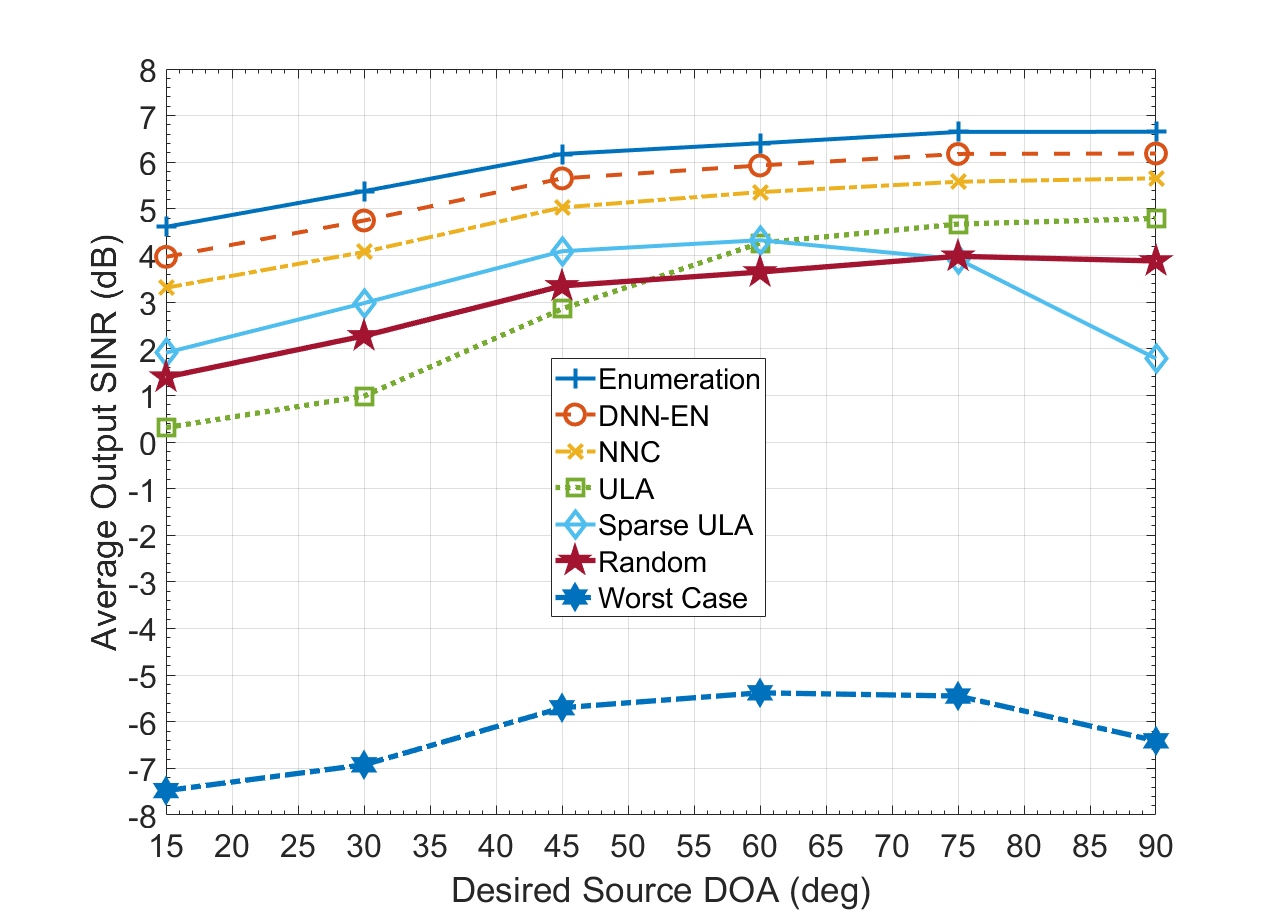}
	\caption{Performance comparison of enumerated design under unlimited snapshots}
	\label{DNN-1}
\end{figure}

Figure \ref{DNN-1} shows the output SINR performance comparisons for different  array configurations.  The horizontal axis is the DOA of the desired point source, and the performance is computed at six different source DOAs varying from   $15^0$ to $90^0$ in  steps of $15^0$.  The SNR of the desired signal is set to $0$ dB, whereas the  interference-to-noise-ratio (INR) for each interference is chosen randomly between $10-20$ dB for a given realization.  The results presented in  Fig. \ref{DNN-1} are obtained by using  unlimited number of data snapshots (USS), i.e., exact spatial correlation values,  and employing  enumerated labels to train the DNN.  This approach is referred to as DNN-EN. The network performance is reported by averaging over  $900$ testing scenarios for each desired source DOA. It is evident that the DNN-EN approach performs  close (0.45 dB trade off) to the performance of the optimum array found by enumeration.  The latter amounts to trying all 980 configurations and choosing the one with the highest SINR. It  involves expensive singular value decomposition (SVD) for each enumeration and is also not scale-able with the problem size, facing the curse of dimensionality. 

The DNN-EN design performance is also compared with the  simple NNC (nearest neighbour correlation) design which returns the label corresponding to the input nearest neighbour correlation function (in terms of mean square error). NNC design is essentially a lookup table, such that for a given test data, it  returns the  label of the closest training example by sifting through the entire training set. It is  noted that the DNN-EN design outperforms the NNC design, which does not involve DNN processing, with an average performance gain of 0.5 dB. The former approach not only offers superior performance but also is more economically viable from `Edge computing' perspective.    It is noted that the nearest neighbour design  requires  maintaining   a large dictionary and run  exhaustive search over the entire  training set  for  each prediction. Similar results are obtained for the NNC  design where  we minimize the   mean absolute error in lieu of  the mean square error. For the underlying case, the DNN-EN approach has around 88$\%$ accuracy for the training data, and has around 54$\%$ accuracy on the test data (meaning that 54$\%$ of the times all $P$ sensor locations are correctly predicted). This is significant since there are $980$ possible permutations for a given example. It is noted that the superior performance of the DNN (shown in Fig. \ref{DNN-1}) is not  simply because it recovered the optimal solution for 54$\%$ of the cases, but also it yielded superior SINR performing configurations for the majority of the remaining 46$\%$ sub optimal solutions. This is    due to the network  ability to generalize the learning as applied to the test set not present in the training data. The superior performance of DNN  over NNC design reveals that  the DNN   doesn't  memorize a lookup table implementation   to locate the nearest training data point and output  the corresponding memorized  optimal sparse configuration.  It is clear from Fig. \ref{DNN-1} that the proposed design yields significant gains over a compact ULA, sparse ULA and randomly selected array topology. The utility of the effective sparse design is also evident from observing the worst case performance which exhibits  significantly low output SINR of around -5 dB on average.

\begin{figure}[!t]
	\centering
	\includegraphics[width=3.6 in, height=2.6 in]{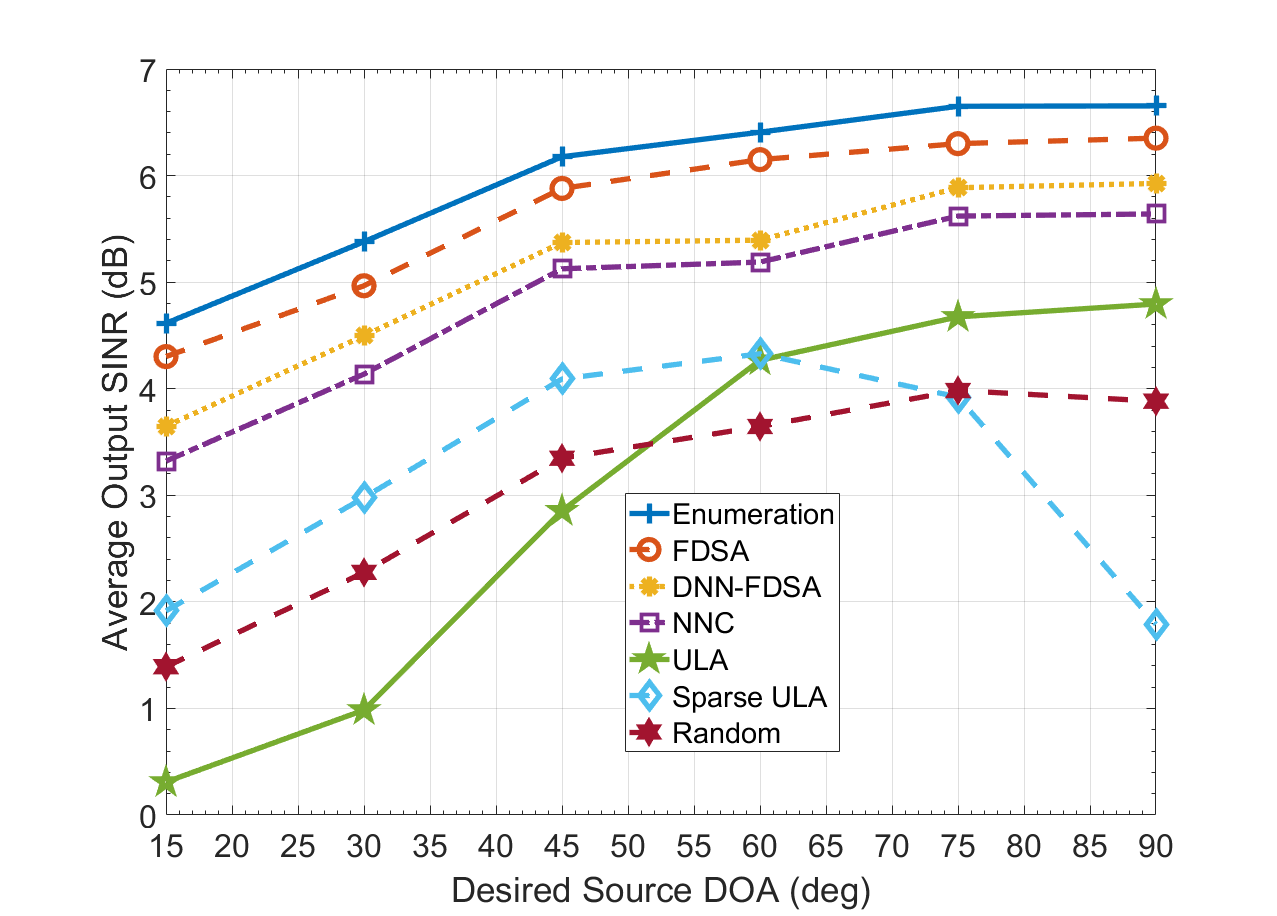}
	\caption{Performance comparison of SBSA design under unlimited snapshots}
	\label{DNN-2}
\end{figure}
\subsection{DNN based SBSA design} \label{Single point source}

Although the training phase is entirely offline, it is infeasible to train the DNN  relying on the enumerated results. This is because  the number of possible sparse solutions can be considerably large even for modest size arrays. For instance, choosing 14 elements out of 24 possible locations results in the order of  $10^6$ candidate spare configurations  for each training data, i.e., for each environment scenario.  In order to avoid this problem and generate a large amount of training  data labels, we resort to  the proposed SBSA design.  Fig. \ref{DNN-2} shows the performance of SBSA design which is merely 0.3 dB down from that of  the design obtained through enumeration.  Quite similar to  the DNN-EN design (0.4 dB down from enumerated  design), the DNN-SBSA design is around 0.38 dB down from SBSA  design. This places the DNN-SBSA design  0.68 dB suboptimal to the enumerated design in aggregate. However, it is still a reasonable performance yielding significant dividends over the commonly used compact ULA, sparse ULA and random sparse topology, as illustrated in the figure.

\subsection{Robust design} \label{Single point source}

\begin{figure}[!t]
	\centering
	\includegraphics[width=3.6 in, height=2.6 in]{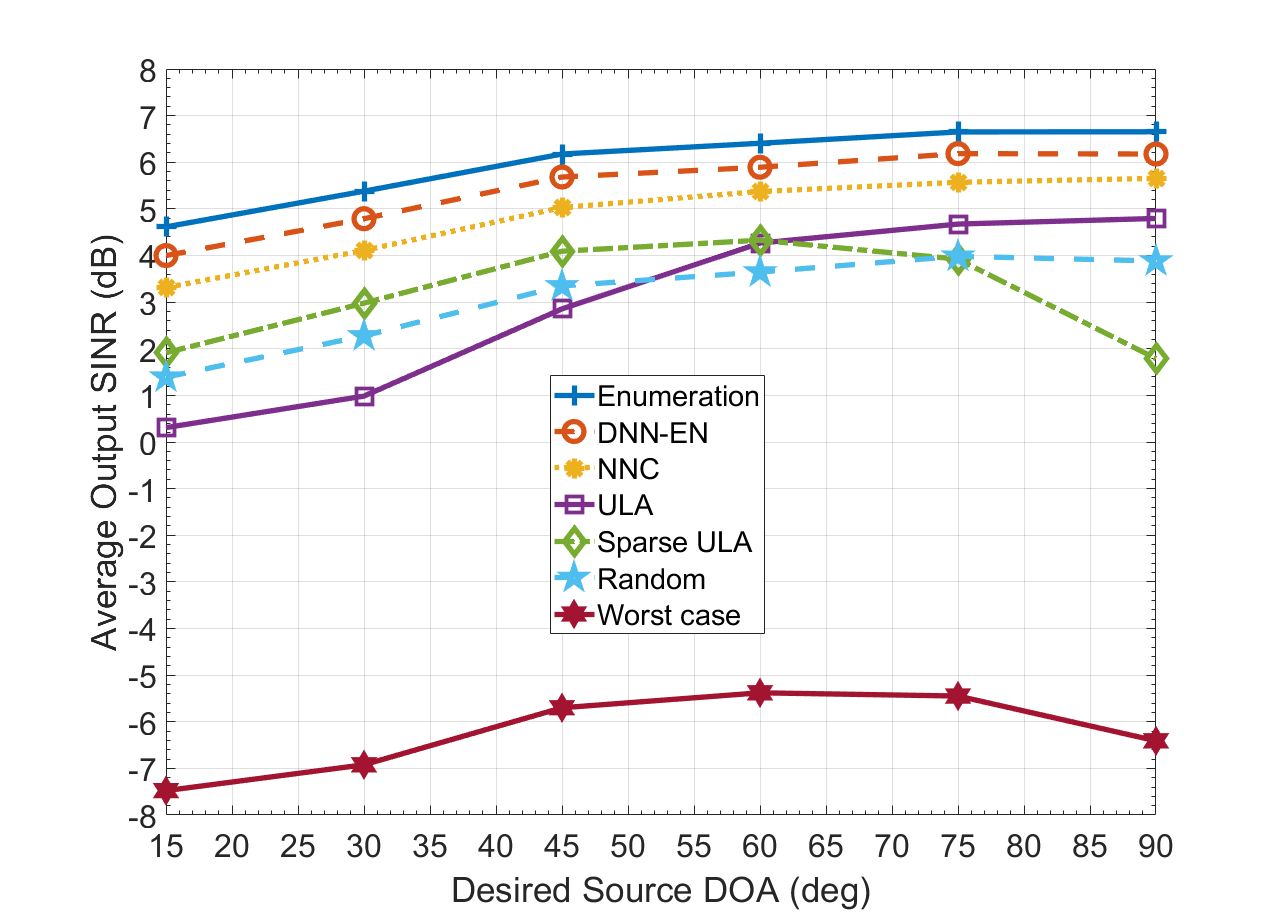}
	\caption{Performance comparison of enumerated design under 512 snapshots}
	\label{DNN-3}
\end{figure}

\begin{figure}[!t]
	\centering
	\includegraphics[width=3.6 in, height=2.6 in]{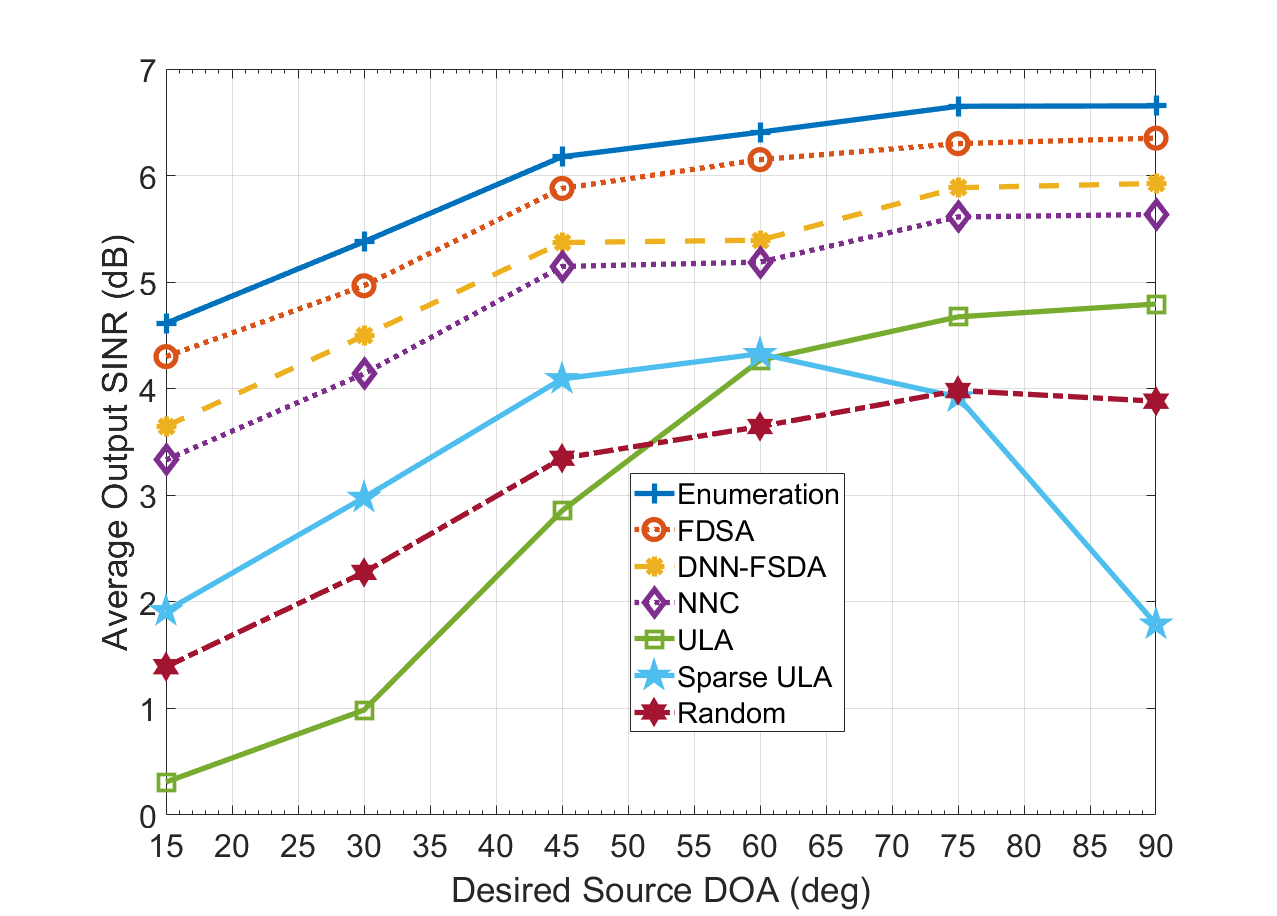}
	\caption{Performance comparison of SBSA design under 512 snapshots}
	\label{DNN-4}
\end{figure}

In order to gauge the robustness of the DNN based scheme, the performance is evaluated under a limited number of data  snapshots. Also, the desired source DOA is perturbed with Gaussian noise of zero mean and 0.25$^0$ variance to account for possible uncertainty around the desired source DOA. For simulating the limited snapshot  scenario, 512 snapshots are generated assuming  the incoming signals (source and interfering signals) are independent BPSK  signals in the presence of Gaussian noise. The correlation matrix under limited snapshots doesn't follow the Toeplitz structure. Therefore, we average the entries along the diagonal and sub-diagonals of the correlation matrix  to calculate the average  values.   Figs. \ref{DNN-3} and  \ref{DNN-4} show the performance of DNN-EN and DNN-SBSA designs under  limited data snapshots. It is clear from the figures that the performance is largely preserved with an SINR discrepancy of less than 0.01 dB demonstrating the robustness of the proposed scheme. The  NNC design, in this case, is suboptimal with more than 0.3 dB additional performance loss.

\subsection{Performance comparisions with state-of-the-art}

\begin{figure}[!t]
	\centering
	\includegraphics[width=3.6 in, height=2.6 in]{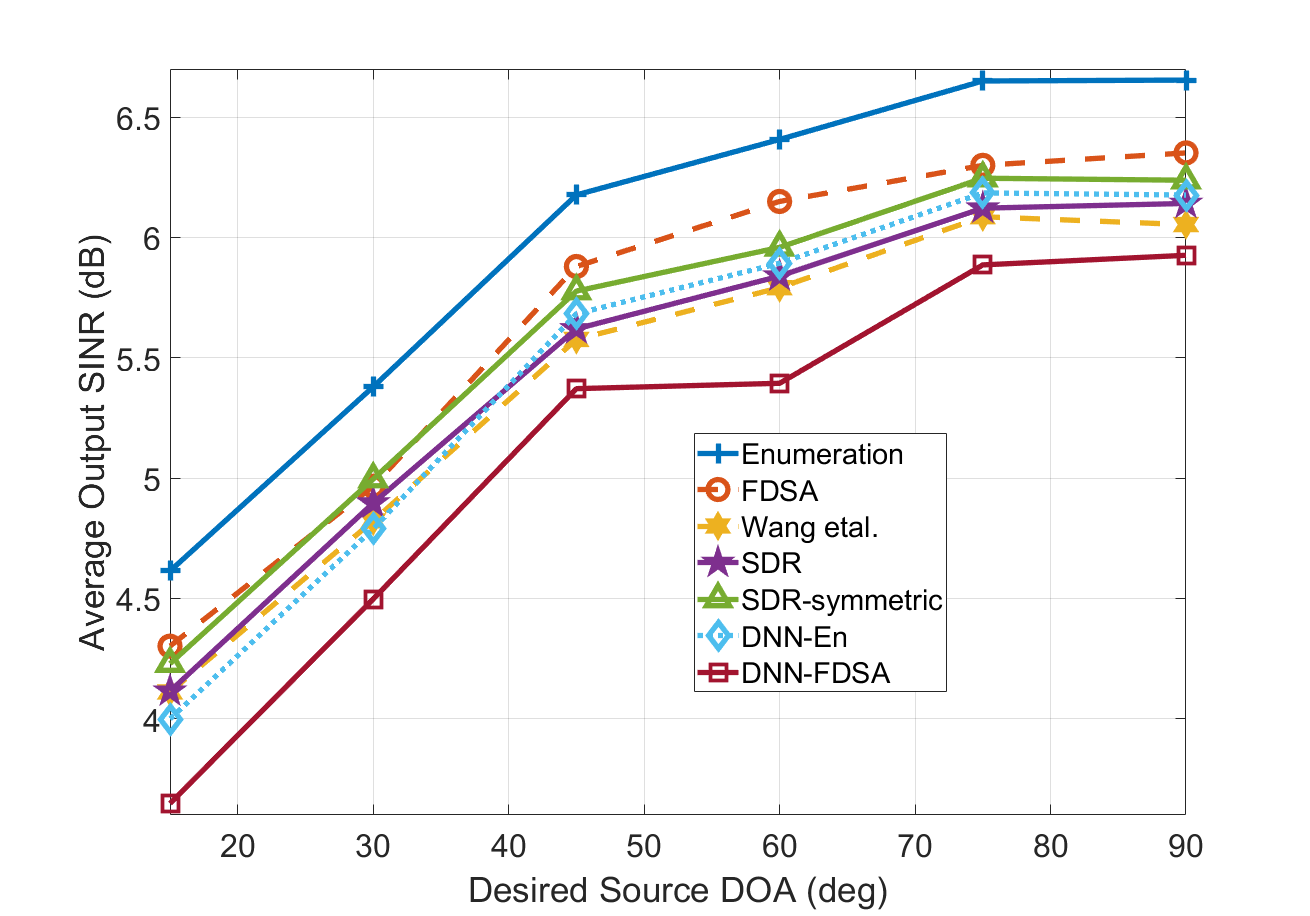}
	\caption{Performance comparisons with the state-of-the-art}
	\label{DNN-5}
\end{figure}

The performance of the proposed SBSA, DNN-EN and DNN-SBSA are compared with  existing work on sparse array design which is based on SDR and SCA approaches \cite{8061016, 8892512}. It is clear from Fig. \ref{DNN-5} that the SBSA algorithm outperforms the other designs and is also more than 100X  computationally efficient as compared to the SDR and SCA (Wang etal. \cite{8061016}) approaches. However, it is only fair to compare the SBSA design with the SCA (Wang etal.) approach because  both incorporate the apriori knowledge of interference parameters. Therefore, in comparing  the data dependent designs, it is found that  SDR design (also the SDR-symmetric \cite{8892512}) is comparable to the  DNN-EN design, with the  DNN-SBSA is marginally suboptimal with the average performance degradation of 0.37 dB.  This slight performance trade off is readily leveraged by the real time realization of the DNN-SBSA algorithm implementing the Capon beamformer in time frames of the order of few milli-seconds.

\label{Single point source}

\section{Conclusion}
This paper considered  sparse array design for maximizing the beamformer output SINR for a desired source in an interference active scenario. A sparse beamformer spectral analysis (SBSA) algorithm was proposed which provided an insightful perspective of the role of array configuration in MaxSINR beamforming. A DNN based approach was developed to configure a data-driven sparse beamformer by learning the enumerated design as well as SBSA design.  We employed  MLP- DNN for its simplicity and limited training data requirements.   Other networks with complex structures, like CNN,  can yield better performance with larger training data set. The proposed methodology harvests  the merits of both the data dependent array designs and  designs assuming prior information of interference parameters. It was shown through  design examples that the  proposed schemes promise optimal solution for around 54$\%$ of the test scenarios and located superior sub-optimal  sparse configurations, in terms of SINR performance,  for the remaining  46$\%$. The proposed approach is robust against limited data snapshots and promise high  performance with reduced computational complexity.

\ifCLASSOPTIONcaptionsoff
  \newpage
\fi
\balance
\bibliographystyle{IEEEtran}
\bibliography{references}

\begin{thebibliography}{10}
\providecommand{\url}[1]{#1}
\csname url@samestyle\endcsname
\providecommand{\newblock}{\relax}
\providecommand{\bibinfo}[2]{#2}
\providecommand{\BIBentrySTDinterwordspacing}{\spaceskip=0pt\relax}
\providecommand{\BIBentryALTinterwordstretchfactor}{4}
\providecommand{\BIBentryALTinterwordspacing}{\spaceskip=\fontdimen2\font plus
\BIBentryALTinterwordstretchfactor\fontdimen3\font minus
  \fontdimen4\font\relax}
\providecommand{\BIBforeignlanguage}[2]{{%
\expandafter\ifx\csname l@#1\endcsname\relax
\typeout{** WARNING: IEEEtran.bst: No hyphenation pattern has been}%
\typeout{** loaded for the language `#1'. Using the pattern for}%
\typeout{** the default language instead.}%
\else
\language=\csname l@#1\endcsname
\fi
#2}}
\providecommand{\BIBdecl}{\relax}
\BIBdecl

\bibitem{710573}
G.~R. Lockwood, J.~R. Talman, and S.~S. Brunke, ``Real-time {3-D} ultrasound
  imaging using sparse synthetic aperture beamforming,'' \emph{IEEE
  Transactions on Ultrasonics, Ferroelectrics, and Frequency Control}, vol.~45,
  no.~4, pp. 980--988, July 1998.

\bibitem{6477161}
O.~Mehanna, N.~D. Sidiropoulos, and G.~B. Giannakis, ``Joint multicast
  beamforming and antenna selection,'' \emph{IEEE Transactions on Signal
  Processing}, vol.~61, no.~10, pp. 2660--2674, May 2013.

\bibitem{1428743}
Y.~He and K.~P. {Chong}, ``Sensor scheduling for target tracking in sensor
  networks,'' in \emph{2004 43rd IEEE Conference on Decision and Control (CDC)
  (IEEE Cat. No.04CH37601)}, vol.~1, Dec 2004, pp. 743--748 Vol.1.

\bibitem{4058261}
W.~V. Cappellen, S.~J. Wijnholds, and J.~D. Bregman, ``Sparse antenna array
  configurations in large aperture synthesis radio telescopes,'' in \emph{2006
  European Radar Conference}, Sept 2006, pp. 76--79.

\bibitem{4663892}
S.~Joshi and S.~Boyd, ``Sensor selection via convex optimization,'' \emph{IEEE
  Transactions on Signal Processing}, vol.~57, no.~2, pp. 451--462, Feb 2009.

\bibitem{6031934}
H.~Godrich, A.~P. Petropulu, and H.~V. Poor, ``Sensor selection in distributed
  multiple-radar architectures for localization: {A} knapsack problem
  formulation,'' \emph{IEEE Transactions on Signal Processing}, vol.~60, no.~1,
  pp. 247--260, Jan 2012.

\bibitem{AMIN20171}
M.~G. Amin, P.~P. Vaidyanathan, Y.~D. Zhang, and P.~Pal, ``Editorial for
  coprime special issue,'' \emph{Digital Signal Processing}, vol.~61, no.
  Supplement C, pp. 1 -- 2, 2017, special Issue on Coprime Sampling and Arrays.

\bibitem{HAMZA2020102678}
\BIBentryALTinterwordspacing
S.~A. Hamza and M.~G. Amin, ``Sparse array design for maximizing the
  signal-to-interference-plus-noise-ratio by matrix completion,'' \emph{Digital
  Signal Processing}, p. 102678, 2020. [Online]. Available:
  \url{http://www.sciencedirect.com/science/article/pii/S1051200420300233}
\BIBentrySTDinterwordspacing

\bibitem{1139138}
A.~Moffet, ``Minimum-redundancy linear arrays,'' \emph{IEEE Transactions on
  Antennas and Propagation}, vol.~16, no.~2, pp. 172--175, March 1968.

\bibitem{5456168}
P.~Pal and P.~P. Vaidyanathan, ``Nested arrays: A novel approach to array
  processing with enhanced degrees of freedom,'' \emph{IEEE Transactions on
  Signal Processing}, vol.~58, no.~8, pp. 4167--4181, Aug. 2010.

\bibitem{7012090}
S.~Qin, Y.~D. Zhang, and M.~G. Amin, ``Generalized coprime array configurations
  for direction-of-arrival estimation,'' \emph{IEEE Transactions on Signal
  Processing}, vol.~63, no.~6, pp. 1377--1390, March 2015.

\bibitem{Goldsmith:2005:WC:993515}
A.~Goldsmith, \emph{Wireless Communications}.\hskip 1em plus 0.5em minus
  0.4em\relax New York, NY, USA: Cambridge University Press, 2005.

\bibitem{Trees:1992:DEM:530789}
H.~L.~V. Trees, \emph{Detection, Estimation, and Modulation Theory: Radar-Sonar
  Signal Processing and Gaussian Signals in Noise}.\hskip 1em plus 0.5em minus
  0.4em\relax Melbourne, FL, USA: Krieger Publishing Co., Inc., 1992.

\bibitem{1206680}
J.~Li, P.~Stoica, and Z.~Wang, ``On robust {Capon} beamforming and diagonal
  loading,'' \emph{IEEE Transactions on Signal Processing}, vol.~51, no.~7, pp.
  1702--1715, July 2003.

\bibitem{6774934}
X.~Wang, E.~Aboutanios, M.~Trinkle, and M.~G. Amin, ``Reconfigurable adaptive
  array beamforming by antenna selection,'' \emph{IEEE Transactions on Signal
  Processing}, vol.~62, no.~9, pp. 2385--2396, May 2014.

\bibitem{1634819}
N.~D. Sidiropoulos, T.~N. Davidson, and Z.-Q. Luo, ``Transmit beamforming for
  physical-layer multicasting,'' \emph{IEEE Transactions on Signal Processing},
  vol.~54, no.~6, pp. 2239--2251, June 2006.

\bibitem{6714077}
V.~Roy, S.~P. Chepuri, and G.~Leus, ``Sparsity-enforcing sensor selection for
  {DOA} estimation,'' in \emph{2013 5th IEEE International Workshop on
  Computational Advances in Multi-Sensor Adaptive Processing (CAMSAP)}, Dec.
  2013, pp. 340--343.

\bibitem{8036237}
X.~Wang, M.~G. Amin, X.~Wang, and X.~Cao, ``Sparse array quiescent beamformer
  design combining adaptive and deterministic constraints,'' \emph{IEEE
  Transactions on Antennas and Propagation}, vol.~PP, no.~99, pp. 1--1, 2017.

\bibitem{8061016}
X.~Wang, M.~Amin, and X.~Cao, ``Analysis and design of optimum sparse array
  configurations for adaptive beamforming,'' \emph{IEEE Transactions on Signal
  Processing}, vol.~PP, no.~99, pp. 1--1, 2017.

\bibitem{8892512}
S.~A. {Hamza} and M.~G. {Amin}, ``Hybrid sparse array beamforming design for
  general rank signal models,'' \emph{IEEE Transactions on Signal Processing},
  vol.~67, no.~24, pp. 6215--6226, Dec 2019.

\bibitem{738234}
J.~{Sheinvald} and M.~{Wax}, ``Direction finding with fewer receivers via
  time-varying preprocessing,'' \emph{IEEE Transactions on Signal Processing},
  vol.~47, no.~1, pp. 2--9, Jan 1999.

\bibitem{967086}
Y.~{Asano}, S.~{Ohshima}, T.~{Harada}, M.~{Ogawa}, and K.~{Nishikawa},
  ``Proposal of millimeter-wave holographic radar with antenna switching,'' in
  \emph{2001 IEEE MTT-S International Microwave Sympsoium Digest (Cat.
  No.01CH37157)}, vol.~2, May 2001, pp. 1111--1114 vol.2.

\bibitem{922993}
{Li Yang}, {Liang Liwan}, {Pan Weifeng}, {Chen Yaqin}, and {Feng Zhenghe},
  ``Signal processing method for switch antenna array of the {FMCW} radar,'' in
  \emph{Proceedings of the 2001 IEEE Radar Conference (Cat. No.01CH37200)}, May
  2001, pp. 289--293.

\bibitem{1299086}
{Moon-Sik Lee}, V.~{Katkovnik}, and {Yong-Hoon Kim}, ``System modeling and
  signal processing for a switch antenna array radar,'' \emph{IEEE Transactions
  on Signal Processing}, vol.~52, no.~6, pp. 1513--1523, June 2004.

\bibitem{articlelecun1}
Y.~LeCun, Y.~Bengio, and G.~Hinton, ``Deep learning,'' \emph{Nature}, vol. 521,
  pp. 436--44, 05 2015.

\bibitem{articlelecun3}
A.~Krizhevsky, I.~Sutskever, and G.~Hinton, ``Imagenet classification with deep
  convolutional neural networks,'' \emph{Neural Information Processing
  Systems}, vol.~25, 01 2012.

\bibitem{inproceedingslecun2}
l.~Deng, J.~Li, J.-T. Huang, K.~Yao, D.~Yu, F.~Seide, M.~Seltzer, G.~Zweig,
  X.~He, J.~Williams, Y.~Gong, and A.~Acero, ``Recent advances in deep learning
  for speech research at microsoft,'' 10 2013, pp. 8604--8608.

\bibitem{DBLP:conf/iclr/LiM17}
K.~Li and J.~Malik, ``Learning to optimize,'' in \emph{5th International
  Conference on Learning Representations, {ICLR} 2017, Toulon, France, April
  24-26, 2017, Conference Track Proceedings}, 2017.

\bibitem{10.5555/3157382.3157543}
M.~Andrychowicz, M.~Denil, S.~G. Colmenarejo, M.~W. Hoffman, D.~Pfau,
  T.~Schaul, B.~Shillingford, and N.~de~Freitas, ``Learning to learn by
  gradient descent by gradient descent,'' in \emph{Proceedings of the 30th
  International Conference on Neural Information Processing Systems}, ser.
  NIPS’16.\hskip 1em plus 0.5em minus 0.4em\relax Red Hook, NY, USA: Curran
  Associates Inc., 2016, p. 3988–3996.

\bibitem{oshea2016recurrent}
T.~J. O'Shea, T.~C. Clancy, and R.~W. McGwier, ``Recurrent neural radio anomaly
  detection,'' \emph{CoRR}, vol. abs/1611.00301, 2016.

\bibitem{8444648}
H.~{Sun}, X.~{Chen}, Q.~{Shi}, M.~{Hong}, X.~{Fu}, and N.~D. {Sidiropoulos},
  ``Learning to optimize: Training deep neural networks for interference
  management,'' \emph{IEEE Transactions on Signal Processing}, vol.~66, no.~20,
  pp. 5438--5453, 2018.

\bibitem{DBLP:journals/twc/ShenSZL20}
Y.~Shen, Y.~Shi, J.~Zhang, and K.~B. Letaief, ``{LORM:} learning to optimize
  for resource management in wireless networks with few training samples,''
  \emph{{IEEE} Trans. Wireless Communications}, vol.~19, no.~1, pp. 665--679,
  2020.

\bibitem{articlesparseli}
P.~Sprechmann, R.~Litman, T.~B. Yakar, A.~M. Bronstein, and G.~Sapiro,
  ``Supervised sparse analysis and synthesis operators,'' in \emph{Advances in
  Neural Information Processing Systems 26: 27th Annual Conference on Neural
  Information Processing Systems 2013, Lake Tahoe, Nevada, United States},
  2013, pp. 908--916.

\bibitem{DBLP:journals/corr/OSheaEC17}
T.~J. O'Shea, T.~Erpek, and T.~C. Clancy, ``Deep learning based {MIMO}
  communications,'' \emph{CoRR}, vol. abs/1707.07980, 2017.

\bibitem{hershey2014deep}
J.~R. Hershey, J.~L. Roux, and F.~Weninger, ``Deep unfolding: Model-based
  inspiration of novel deep architectures,'' \emph{CoRR}, vol. abs/1409.2574,
  2014.

\bibitem{doi:10.1137/080716542}
A.~Beck and M.~Teboulle, ``A fast iterative shrinkage-thresholding algorithm
  for linear inverse problems,'' \emph{SIAM Journal on Imaging Sciences},
  vol.~2, no.~1, pp. 183--202, 2009.

\bibitem{10.5555/3104322.3104374}
K.~Gregor and Y.~LeCun, ``Learning fast approximations of sparse coding,'' in
  \emph{Proceedings of the 27th International Conference on International
  Conference on Machine Learning}, ser. ICML’10.\hskip 1em plus 0.5em minus
  0.4em\relax Madison, WI, USA: Omnipress, 2010, p. 399–406.

\bibitem{elbir2019cognitive}
A.~Elbir, K.~V. Mishra, and Y.~Eldar, ``Cognitive radar antenna selection via
  deep learning,'' \emph{IET Radar, Sonar \& Navigation}, vol.~13, 02 2018.

\bibitem{elbir2019joint}
A.~Elbir and K.~Mishra, ``Joint antenna selection and hybrid beamformer design
  using unquantized and quantized deep learning networks,'' \emph{IEEE
  Transactions on Wireless Communications}, vol.~19, pp. 1677--1688, 2020.

\bibitem{elbir2019joint2}
A.~Elbir, K.~V. Mishra, and B.~S. Mysore, ``Online and offline deep learning
  strategies for channel estimation and hybrid beamforming in multi-carrier
  mm-wave massive mimo systems,'' \emph{arXiv: Signal Processing}, 07 2020.

\bibitem{9108299}
A.~M. Elbir and K.~V. Mishra, ``Sparse array selection across arbitrary sensor
  geometries with deep transfer learning,'' \emph{IEEE Transactions on
  Cognitive Communications and Networking}, vol.~7, no.~1, pp. 255--264, 2021.

\bibitem{trove.nla.gov.au/work/15617720}
P.~Stoica and R.~L. Moses, \emph{\BIBforeignlanguage{English}{Introduction to
  spectral analysis}}.\hskip 1em plus 0.5em minus 0.4em\relax Upper Saddle
  River, N.J. : Prentice Hall, 1997.

\bibitem{1223538}
S.~Shahbazpanahi, A.~B. Gershman, Z.-Q. Luo, and K.~M. Wong, ``Robust adaptive
  beamforming for general-rank signal models,'' \emph{IEEE Transactions on
  Signal Processing}, vol.~51, no.~9, pp. 2257--2269, Sept. 2003.

\bibitem{DBLP:journals/corr/KingmaB14}
D.~P. Kingma and J.~Ba, ``Adam: {A} method for stochastic optimization,'' in
  \emph{3rd International Conference on Learning Representations, {ICLR} 2015,
  San Diego, CA, USA, May 7-9, 2015, Conference Track Proceedings}, Y.~Bengio
  and Y.~LeCun, Eds., 2015.

\bibitem{554195}
S.~Lawrence, C.~Giles, A.~C. Tsoi, and A.~Back, ``Face recognition: a
  convolutional neural-network approach,'' \emph{IEEE Transactions on Neural
  Networks}, vol.~8, no.~1, pp. 98--113, 1997.

\bibitem{GU2018354}
\BIBentryALTinterwordspacing
J.~Gu, Z.~Wang, J.~Kuen, L.~Ma, A.~Shahroudy, B.~Shuai, T.~Liu, X.~Wang,
  G.~Wang, J.~Cai, and T.~Chen, ``Recent advances in convolutional neural
  networks,'' \emph{Pattern Recognition}, vol.~77, pp. 354--377, 2018.
  [Online]. Available:
  \url{https://www.sciencedirect.com/science/article/pii/S0031320317304120}
\BIBentrySTDinterwordspacing

\end{thebibliography}
\end{document}